\documentclass[10pt,twocolumn]{IEEEtran}

\usepackage[ruled]{algorithm2e}
\usepackage{emptypage}

\usepackage{amsthm}
\usepackage{enumerate}
\usepackage{mathrsfs}
\usepackage{amssymb}
\usepackage{graphicx}
\usepackage{multirow}
\usepackage[cmex10]{amsmath}
\usepackage{subfigure}
\usepackage{diagbox}
\usepackage{cite}
\usepackage{color}
\usepackage{chngpage}
\usepackage{algpseudocode}
\usepackage{amsmath}

\allowdisplaybreaks

\begin{document}

\makeatletter
\renewcommand{\maketag@@@}[1]{\hbox{\m@th\normalsize\normalfont#1}}%
\makeatother
\title{Uplink Channel Estimation and Signal Extraction Against Malicious IRS in Massive MIMO System }
\pagestyle{empty}
\author{Xiaofeng Zheng, Ruohan Cao, and Lidong Ma
 \thanks{X. Zheng, R. Cao and L. Ma are with the Key Laboratory of Trustworthy Distributed Computing and Service, Ministry of Education, and also the School of Information and Communication Engineering, Beijing University of Posts and Telecommunications (BUPT), Beijing 100876, China (e-mail:\{zhengxiaofeng, caoruohan, mald2020\}@bupt.edu.cn).}}
\maketitle
\pagestyle{empty}
\begin{abstract}
This paper investigates effect of malicious intelligence reflecting surface (IRS). The malicious IRS is utilized for performing attack by randomly reflecting data sequences of legitimate users (LUs) to a base station (BS). We find that the data sequences of LUs are correlative to the signals reflected by malicious IRS. The correlation undermines the performance of traditional eigenvalue decomposition (EVD)-based channel estimation (CE) methods.\ To address this challenge, we propose a empirical-distribution-based channel estimation approach in the presence of malicious IRS. The proposed method works by capturing desired convex hulls from signals disturbed by malicious IRS, on the basis of its empirical distribution. Simulation results show that our proposed approach outperforms traditional EVD-based methods as much as nearly 5 dB in normalized mean square error (NMSE).
\end{abstract}

\IEEEpeerreviewmaketitle
\begin{IEEEkeywords}
Malicious attack, uplink channel estimation, massive MIMO,  intelligent reflecting surface
\end{IEEEkeywords}
\section{Introduction}
Massive multiple-input and multiple-output (MIMO) is a  key technology in fifth-generation (5G) communication that can achieve high speed and large capacity \cite{5Gwhole}. Its advantage  depends  on trustworthy channel state information (CSI) \cite {twoway}.  However, malicious users (MUs) may exist in 5G networks, and they may actively send interference  to disturb channel estimation.  As a result, trustworthy CSI cannot be obtained, and false CSI undermines {\color{black}the} performance of a massive MIMO system. To obtain trustworthy CSI,\ it is important to investigate channel estimation under malicious attack \cite{5gsecurity}.
\par
Much work has investigated channel estimation under malicious attack. In\cite {receiver}, legitimate user (LU) and base station (BS) share a secret PS that is unknown to MU. The secret pilot sequence (PS) then enables channel estimation. In\cite{LUknow1}, all LUs and MUs select random PSs from a well-known pilot codebook that consists of orthogonal PSs. Based on the codebook, the BS firstly estimates the selected PSs, and then estimate channels. In\cite {ica1}, LUs and MUs independently send random symbols. Based on the independence, independent component analysis (ICA) could be invoked for channel estimation.

The above methods are implemented during the pilot phase.  Other works estimate channels by employing eigenvalue decomposition (EVD) to signals received through the data phase. Based on the resulting eigenspaces, channels of LUs and MUs can be separated in probability as the number of antennas approaches infinity \cite{evd2}\cite{doubletrain}. The transmission power gap between LUs and MUs is  assumed and  used for channel identification.

The works above assume that the MUs are equipped with traditional transmitters. On the other hand, intelligent reflecting surface (IRS), as a promising device, has attracted much attention. When signals propagate to IRS, the IRS could reflect the signals with programmable phase adjustment. By properly reflecting signals according to predesignated phase adjustment protocols, IRS could cooperate on  channels estimation\cite{9133142}, or enlarging secrecy rate \cite{irssecure1}\cite{irssecure2}. It is worth noting that  IRS is only assumed to work as collaborator in prior works. However, to authors' best knowledge, it is sparse to consider that IRS is used for attack.

In this paper, we consider malicious IRS. The IRS reflects pilot or data signals from LUs with unknown phase adjustment. The reflection signals propagate to BS, and interfere signal reception at the BS. Since the IRS is different with traditional active transmitters, existing methods based on active transmitter may be not applicable to combat malicious IRS \cite{receiver, LUknow1, ica1, evd2,doubletrain}. To this problem, we propose a channel estimation and signal extraction method in the presence of malicious IRS.

The main contributions of the paper are  as follows{\color{black}{.}}
{\color{black}{\begin{enumerate}
\item We find that there is correlation between the reflecting signal and the legitimate signal.  And the correlation degrades the performance of traditional channel estimation methods based on EVD of the received signals\cite{evd2}\cite{doubletrain}.
\item To combat the attacks caused by the malicious IRSs, we use a geometric argument to develop signal extraction and channel estimation criteria. The geometric argument is robust to attack, but sensitive to noise. To optimize the proposed criteria, we presents an extractor to obtain geometric properties of desired signals from noisy observations. With the help of the extractor, we achieve signal extraction and channel estimation in the presence of attacks by solving two optimization problems.

%%%\item Firstly, we consider to extract data symbols from noisy received signals under attack, as a basis of channel estimation. Due to the possible existence of \emph{correlative} attack, we utilize desired signals' (DSs) geometry property, which exhibits robustness to correlation attack. DS is the product of channels and random data/interference without noise. To be more precise, the symbol extraction corresponds with the minimization of convex perimeter of DSs' linear combination (LC-DS). Notice that we only get noisy observation of LC-DS, rather than  LC-DS itself. And the convex perimeter is sensitive to noise. Hence, we propose an extractor that can distill the convex perimeter against attack and noise. As a beneficial result, data symbols are extracted in the presence of attack.
%The extractor relies on the fact  that the CF of noisy observations is the product of the CFs of  TS and noise. Motivated by this, we formulate an optimization problem, the objective is to get the distribution of TS, then according to this distribution to distill alphabets.
%\item Secondly, on the basis of symbol extraction, we estimate the channel corresponding to the extracted symbols.
\thispagestyle{empty}
%%%%Notice that the channel coefficient corresponds with the minimization of convex perimeter of DSs with a deduction of weighted extracted symbols. This observation motivates us to formulate an optimization problem, while refine possible solutions into a finite and discrete set. Both the optimization problem and the solution set are established relying on our proposed extractor again.According to the finite and discrete solution set, we find the global optimum solution, and achieve channel estimation in the presence of attack.
% To estimate the channel,  we prove that the possible channel alphabets  of every antenna are in a finite set, the real channel alphabet  let the perimeter of signal separation  minimum. Further we use the extractor to get the alphabets of every row of observation and extracted signal, then get the finite set. Choose alphabets from the finite set, which let the perimeter of signal separation  minimum.  Finally, we get the channel estimation.
%We find that the alphabets of DS under the influence of channel. Further, we can estimate the channel according to this. The noise influence the alphabets of desired signal, we can use extractor to get the clean alphabets. Then, get the channel estimation.
\end{enumerate}}}

\emph{Notation:} Vectors are denoted by lowercase italicized letters, and matrices  by uppercase italicized letters. A superscript $(\cdot)^T$ indicates a matrix transpose. We use  tr($\textbf{\emph{A}}$) to denote the trace of matrix $\textbf{\emph{A}}$, and $[\cdot]_m$ denotes the $m$th row of an input matrix or vector.   $P_X$ denotes the {stochastic} distribution of the random variable $X$, $P_X(x)=\Pr(X=x)$.   $P_{AB}$ denotes the joint distribution of  random variables $A$ and $B$.
$\odot$ denotes a dot product. $ X \xrightarrow{a.s.} Y$ indicates that $X$ converges to $Y$ almost surely, where $X$ and $Y$ are generic random variables or bounded constants.
$\left\|\cdot \right\|_2$ denotes the 2-norm.

\clearpage
\label{sec:intro}
\section{ System Model} \label{system model}
{\color{black}In~Fig.~\ref{fig:attackmodel}, we consider system model including a BS equipped with $M$ antennas, {\color{black}$N$} single-antenna LUs, and $N$ MUs,
where the $j$th LU is attacked by {\color{black} the $j$th MU}, $j=1,2,\cdots, N$.
%\footnote{This assumption also follows ref.\cite{onemu}.}   
Each MU is equipped with an IRS that includes $W$ elements.}
The uplink communication between the BS and the LUs takes place in the pilot and data phases, including $L_p$ and $n$
instants, respectively. The $j$th MU is assumed to locate close to the $j$th LU, and far away from other LUs. We thus assume that the $j$th MU conducts attack by only reflecting signals from the $j$th LU in the two phases. During the pilot phase, the IRSs of MUs reflect PSs without any phase-shift, and during the data phase, the MUs reflect information data sequences with random phase shift.
%The \emph{correlative} attack model and its effects are detailed below.
%interference to the BS.
%The BS estimates the uplink channels in the presence of attack by based on its observation.
\thispagestyle{empty}

%Let us consider there is $N$ LUs and $N$ MUs, in this attack model,
{\color{black}To be more precisely, in the pilot phase, the $j$th LU transmits PS $\emph{\textbf{x}}_j\in \mathbb{C}^{1\times L_p}$, which  is selected from a public or secret pilot codebook $\emph{\textbf{X}}$. The $j$th MU conducts pilot spoof attack, by using a identity matrix $\Phi_{p}= diag(1,\cdots,1)$, $\Phi_{p}\in \mathbb{C}^{W\times W}$, as reflection-coefficient matrix of its IRS. In this way, the MU reflects $\emph{\textbf{x}}_j$ without any phase adjustment. Then, the reflection signal is same to $\emph{\textbf{x}}_j$, which constitutes pilot spoof attack even the pilot codebook $\emph{\textbf{X}}$ is
unknown to the MUs.}

The received signals $\textbf{\emph{Y}}_p \in\mathbb{C}^{M\times L_p}$  in the pilot phase  can be   specified as {\color{black}\begin{align} \footnotesize 
\textbf{\emph{Y}}_p=\nonumber & \footnotesize {\sum_{j=1}^{N}}(\emph{\textbf{h}}_j\emph{\textbf{x}}_j+\emph{\textbf{G}}_{j_2}\Phi_{p}\emph{\textbf{g}}_{j_1}\emph{\textbf{x}}_j) + \textbf{\emph{N}}_p\\=& \footnotesize {\sum_{j=1}^{N}}(\emph{\textbf{h}}_j\emph{\textbf{x}}_j+{\sum_{w=1}^{W}}\emph{\textbf{g}}_{j_w}\emph{\textbf{x}}_j) + \textbf{\emph{N}}_p,
\end{align}
where $\emph{\textbf{h}}_j$ denotes the channel from the $j$th LU to the BS,  $\emph{\textbf{h}}_j\in\mathbb{C}^{M\times 1}$; $\emph{\textbf{g}}_{j_1}, \emph{\textbf{G}}_{j_2}$ respectively denote the channels from the $j$th LU to the $j$th MU and  from the $j$th MU to the BS, $\emph{\textbf{g}}_{j_1}=[\emph{{g}}_{j_1}(1),\cdots,\emph{{g}}_{j_1}(W)]^T\in\mathbb{C}^{W\times 1}$, $\emph{{g}}_{j_1}(w) \in\mathbb{C}^{1\times 1}$,  $\emph{{g}}_{j_1}(w)$ denotes the channel from the $j$th LU to the $w$th elememt of the $j$th MU.  $\emph{\textbf{G}}_{j_2}=[\emph{\textbf{g}}_{j_2}[1], \cdots,\emph{\textbf{g}}_{j_2}[W]] \in\mathbb{C}^{M\times W}$, $\emph{\textbf{g}}_{j_2}[w]\in\mathbb{C}^{M\times 1}$, $\emph{\textbf{g}}_{j_2}[w]$ denotes the channel from the $w$th elements of $j$th MU to the BS; and $\emph{\textbf{g}}_{j_w}=\emph{\textbf{g}}_{j_2}[w]\emph{{g}}_{j_1}(w)$,  denotes the cascaded channels of the $w$th element of IRS, $\emph{\textbf{g}}_{j_w}\in\mathbb{C}^{M\times 1}$.
 $\textbf{\emph{N}}_p\in\mathbb{C}^{M\times L_p}$  are Gaussian noise, and each element follows $\mathcal{CN}(0, \sigma^{2})$.

In the data phase, the $j$th LU transmits $ \emph{\textbf{a}}_j \in\mathbb{C}^{1\times n}$. Due to the  IRS of MU with $W$ reflection elements, the MU reflects $W$ stream signal sequences. We further define the diagonal matrix $\Phi_j(t)= diag(e^{\mathsf{i}\phi_{j_1}(t)},\cdots,e^{\mathsf{i}\phi_{j_W}(t)})$, $1\leq t\leq n$, $\Phi_{j}(t)\in \mathbb{C}^{W\times W}$ as the reflection-coefficient
matrix of the  $j$th  IRS,  which is randomly set according to $\Pr\{\phi_{j_w}(t)=0\}=p_w$, $\Pr\{\phi_{j_w}(t)=\pi\}=1-p_w, 1\leq w\leq W$}. The received signals $\textbf{\emph{y}}(t) \in\mathbb{C}^{M\times1}$  in the data phase  can be   specified as
 {\color{black}
\begin{align} \footnotesize \label{rece}
\textbf{\emph{y}}(t)& \nonumber= \footnotesize{ \sqrt{P} {\sum_{j=1}^{N}}(\emph{\textbf{h}}_j{\emph{\textbf{a}}_j}(t)+\emph{\textbf{G}}_{j_2}\Phi_j(t)\emph{\textbf{g}}_{j_1}{\emph{\textbf{a}}_j}(t))+\textbf{\emph{N}}}\\&= \footnotesize \sqrt{P} {\sum_{j=1}^{N}}(\emph{\textbf{h}}_j{\emph{\textbf{a}}_j}(t)+{\sum_{w=1}^{W}}\emph{\textbf{g}}_{j_w}{\emph{\textbf{b}}_{j_w}}(t))+\textbf{\emph{n}},1\leq t\leq n
\end{align}
\begin{equation}\label{coree}
{ \footnotesize \emph{\textbf{b}}_{j_w}}(t)={\emph{\textbf{a}}_j}(t)e^{\mathsf{i}\phi_{j_w}(t)},
\end{equation}
}{\color{black}$\emph{\textbf{b}}_{j_w}  \in\mathbb{C}^{1\times n}$, where ${\emph{\textbf{a}}_j}(t), {\emph{\textbf{b}}_{j_w}}(t)$ respectively denote the $t$th elements in ${\emph{\textbf{a}}_j}, {\emph{\textbf{b}}_{j_w}}$.  $\textbf{\emph{n}} \in\mathbb{C}^{M\times 1}$ is Gaussian noise, and each element follows $\mathcal{CN}(0, \sigma^{2})$. $P$ is the transmission power.{

{\color{black}
\emph{Remark 1:}
Although we consider each LU to be attacked by single MU with $W$ elements,  the model characterized by (\ref{rece}) and (\ref{coree}) is equivalent the two-MU  with
each  having $\frac{W}{2}$ reflection elements. As such, our proposed technique is  extensible for multi-MU model.
}
\section{Attack Strategies And Effects}\label{attack}
\subsection{Attack strategies}
{\color{black}
The malicious IRS may perform deterministic and random reflection. These strategies are characterized by $p_w$. The deterministic reflection corresponds with $p_w=1$ or $p_w=0$. In other words, the IRSs reflect the signals of LUs with same or opposite phase.  Then, the conventional EVD-based methods can be used to estimate composite channels
${\emph{\textbf{h}}_j}\pm{\sum_{w=1}^{W}}{\emph{\textbf{g}}_{j_w}}$, ${\emph{\textbf{a}}_j}$ could be decoded based on channel estimation.

The main challenge is brought by  the random reflection, wherein  $0< p_w < 1$.
 The random reflection causes correlation between $\emph{\textbf{b}}_{j_w}$ and $\emph{\textbf{a}}_j$. To define this attack strategy mathematically, we assume that $\emph{\textbf{a}}_j$ is an independent and identically distributed (i.i.d.) sequence. According to (\ref{coree}),  ${\emph{\textbf{b}}_{j_w}}$ is also an i.i.d. sequence. There are random variables $A$ and $B$ having the same
stochastic distributions as   each element of $\emph{\textbf{a}}_j$ and $\emph{\textbf{b}}_{j_w}$, respectively.
Let us use $P_A$ and $P_B$ to denote
stochastic distributions of $A$ and $B$, respectively. $\mathcal{A}_j$ and $\mathcal{B}_{j_w}$ are the alphabets of these two variables, $\mathsf{a}$ and $\mathsf{b}$ denote generic symbols of $\mathcal{A}_j$ and $\mathcal{B}_{j_w}$, respectively.  When BPSK modulation is used by the LUs,
it is not hard to obtain that  $P_{A}(1)=P_{A}(-1)=\frac{1}{2}$,  $P_{B|A}(1|1)=P_{B|A}(-1|-1)=p_w$, $P_{B|A}(-1|1)=P_{B|A}(1|-1)=1-p_w$. Therefore,
\begin{equation} \footnotesize \label{dis1}
P_{A,B}(1,1)
=\frac{1}{2}p_w, P_{A}(1)P_{B}(1)=\frac{1}{4}.
\end{equation}
By designing  $p_w\neq \frac{1}{2}$ in (\ref{coree}),  there exists
%Then the statistic definition of \emph{correlative} attack is given by:
}
\begin{equation} \footnotesize \label{occur}
 P_{A,B}({\mathsf{a},\mathsf{b}}) \neq P_{A}({\mathsf{a}})P_{B}({\mathsf{b}}), \mathsf{a}\in \mathcal{A}_j, \mathsf{b}\in \mathcal{B}_{j_w}.
 \end{equation}
\thispagestyle{empty}{\color{black}Eq. (\ref{occur}) shows  that ${\emph{\textbf{a}}_j}$ and ${\emph{\textbf{b}}_{j_w}}$  are \emph{correlative},  and we  refer the attack characterized by (\ref{occur}) as a \emph{correlative} attack. We further find  that\footnote{More detailed proof is presented in Appendix A.} the correlation coefficient between $\emph{\textbf{a}}_j$ and $\emph{\textbf{b}}_{jw}$ is given by $2p_w-1$. This indicates that the  MU can control
the strength of a \emph{correlative} attack by adjusting its reflection probability $p_w$\footnote{When $p_w=\frac{1}{2}$, $P_{A,B}(1,1) = P_{A}(1)P_{B}(1)$,  then ${\emph{\textbf{a}}_j}$ and $ {\emph{\textbf{b}}_{j_w}}$ are independent.  EVD-based methods can be used to estimate channels based on independent data sequences\cite{evd2}\cite{doubletrain}}.

%When $p_w=\frac{1}{2}$, $P_{A,B}(1,1) = P_{A}(1)P_{B}(1)$,  then ${\emph{\textbf{a}}_j}$ and $ {\emph{\textbf{b}}_{j_w}}$ are independent, EVD-based methods can be used to estimate channels.

 In summary, the MUs do not need to explicitly know $\emph{\textbf{a}}_j$. By setting $p_w\neq\frac{1}{2}$, \emph{correlative} attack can be conducted.
% It is worth noting that IRS can be flexibly deployed in wireless networks \cite{magine} to achieve  high spectrum and energy efficiency \cite{energy}.
%to improve their performance \cite{magine}, such as improving  the physical layer security by setting the IRS in the vicinity of the MU to cancel out the signal from the MU to BS,  deploying  the IRS at the cell edge to help improve the desired signal power, setting the IRS then get a link for users in a dead zone where the direct link between it and its serving BS is severely blocked by an obstacle.
%Constrained by locations, some IoT devices may possess unreliable communication channels to the BSs due to the obstacles between them, to tackle this challenge, employ  IRS for enhancing the strength of the IoT devices’ signals at the BSs,
%Thus,  IRSs have the potential of  wide use by the IoT to enhance its coverage and efficiency \cite{iot2}.
%Nevertheless, to the authors' best knowledge, there has been little investigation of the detriment of IRSs employed by MUs. The  example illustrates that  IRSs can be used to perform \emph{correlative} attacks.
We  proceed to analyze its effect below.

%Every element in ${\Phi}\in\mathbb{C}^{1\times n}$ follows $e^{j\phi}$, where $\phi\in\{0,\pi\}$, the probability of selecting $0$ and $\pi$ are different, are $p$ and $1-p$, respectively. Then the correlation coefficient of $\emph{\textbf{a}}_j$ and $\emph{\textbf{b}}_k$ is $2p-1$.  Due to space limitation,  we omit the proof.
%MUs select the $p\neq\frac{1}{2}$, the $\emph{\textbf{a}}_j$ and $\emph{\textbf{b}}_k$ have correlation. That means,  in this scenario, there have \emph{correlative} attack.

\subsection{Detriment of  Correlative Attack}
{\color{black}In the pilot phase, after receiving $\textbf{\emph{Y}}_p$, the BS may {\color{black}estimate} channels of the LUs by projecting $\textbf{\emph{Y}}_p$ onto $\emph{\textbf{X}}$,
\begin{equation} \footnotesize \label{estH}
\begin{split}
&\left[\tilde{\emph{\textbf{h}}}_1,\cdots,\tilde{\emph{\textbf{h}}}_{N}\right] =  \textbf{\emph{Y}}_p\emph{\textbf{X}}^H \\ &=\left[{\emph{\textbf{h}}}_1,\cdots,{\emph{\textbf{h}}}_{N}\right]+ \left[{\emph{\textbf{g}}}_{1_1},\cdots,{\emph{\textbf{g}}}_{1_W}, \cdots, {\emph{\textbf{g}}}_{N_1} \cdots,{\emph{\textbf{g}}}_{N_W}  \right]
  +\textbf{\emph{N}}_p\emph{\textbf{X}}^H,
\end{split}
\end{equation}}where the second equality  relies on the orthogonal property of $\emph{\textbf{X}}$.
This  indicates that the pilot spoof attack causes the channel estimation to combine the legitimate and malicious channels. There is a large estimation error. It is difficult to obtain trustworthy CSI only using $\textbf{\emph{Y}}_p$. We propose to use $\textbf{\emph{Y}}$ received during the data phase for channel estimation.
%In order to obtain more accurate channel estimation, the BS may utilize $\textbf{\emph{Y}}$ in DP.

By collecting  all the $\textbf{\emph{y}}(t)$, $1\leq t\leq n$  in a transmission block, the received signal  in the data phase  can be recast as
\begin{align} \footnotesize \label{matrixform}
\textbf{\emph{Y}}
%&={\sum_{j=1}^{N}}\textbf{\emph{C}}_j\textbf{\emph{S}}_j +\textbf{\emph{N}} \\
=\textbf{\emph{C}} \textbf{\emph{S}} +\textbf{\emph{N}},
\end{align}
{\color{black}where $\footnotesize\textbf{\emph{C}}=\left[ \textbf{\emph{H}} ,\textbf{\emph{G}}\right] $, $\footnotesize\textbf{\emph{S}}=\left[ \begin{matrix}  \textbf{\emph{A}},  \textbf{\emph{B}} \end{matrix} \right]$,
$\footnotesize\textbf{\emph{Y}}$$=$$\left[\textbf{\emph{y}}(1),\textbf{\emph{y}}(2),\cdots,\textbf{\emph{y}}(n)\right]$,
\\$\footnotesize\textbf{\emph{H}}=[\emph{\textbf{h}}_1,\emph{\textbf{h}}_2,\cdots,\emph{\textbf{h}}_{N}]$,  $\footnotesize{\textbf{\emph{A}}=\sqrt{P} \left[\begin{matrix}\emph{\textbf{a}}^T_1,\emph{\textbf{a}}^T_2,\cdots,\emph{\textbf{a}}^T_{N}\end{matrix}\right]^T}$,\\
$\footnotesize\textbf{\emph{G}}$$=$$\left[{\emph{\textbf{g}}}_{1_1},\cdots,{\emph{\textbf{g}}}_{1_W}, \cdots, {\emph{\textbf{g}}}_{N_1}, \cdots,{\emph{\textbf{g}}}_{N_W}  \right]$,
$\footnotesize{\textbf{\emph{B}}=\sqrt{P} \left[ \begin{matrix} \emph{\textbf{b}}^T_{1_1} , \cdots,\emph{\textbf{b}}^T_{1_W}, \cdots,  \emph{\textbf{b}}^T_{{N_1}}, \cdots,  \emph{\textbf{b}}^T_{{N_W}}\end{matrix}\right]^T}$. $\footnotesize\textbf{\emph{Y}} \in\mathbb{C}^{M\times n}$, and each element in  $\textbf{\emph{N}} \in\mathbb{C}^{M\times n}$ follows $\mathcal{CN}(0, \sigma^{2})$. }

Traditional methods apply EVD to $\frac{1}{Mn}{\emph{\textbf{{Y}}}}{\emph{\textbf{{Y}}}}^H$. The resulting eigenspace is then used for jamming rejection when the jamming and legitimate data sequences are independent, i.e., $\frac{\emph{\textbf{{S}}}\emph{\textbf{{S}}}^H}{n}\xrightarrow{a.s.}{{\emph{\textbf{I}}_{N+WN}}}$.  However, under \emph{correlative} attacks, due to (\ref{occur}), we have
$\frac{\emph{\textbf{{S}}}\emph{\textbf{{S}}}^H}{n}\xrightarrow{a.s.}R_s \in\mathbb{C}^{(N+WN)\times (N+WN)}$, where $R_s\neq {{\emph{\textbf{I}}_{N+WN}}}$. We find that a \emph{correlative} attack undermines the performance of an
EVD-based method  using the received signal $\textbf{\emph{Y}}$ \cite{evd2}\cite{doubletrain}.
\thispagestyle{empty}
\newtheorem*{mypro3}{Proposition 1}
\begin{mypro3}
{\color{black}In a large-scale antenna regime, the right singular matrix of   $\frac{1}{Mn}{\textbf{{Y}}}{\textbf{{Y}}}^H$  is $U_{\textbf{{Y}}}=\frac{1}{\sqrt{M}} [U_W, \textbf{{C}}{{R}}^{-\frac{1}{2}}U_s]$, where $U_W  \in\mathbb{C}^{M\times (M-N-WN)}$ has orthogonal columns and spans the null space of $[\textbf{{C}}{{R}}^{-\frac{1}{2}}U_s]$, $R\in\mathbb{C}^{(N+WN)\times (N+WN)}$ is a diagonal matrix depending on $\textbf{{C}}$, $U_s\in\mathbb{C}^{(N+WN)\times (N+WN)}$ is an orthogonal matrix, $\Lambda\in\mathbb{C}^{(N+WN)\times (N+WN)}$ is a diagonal matrix, and they are results of eigenvalue decomposition, i.e.,  $R^{\frac{1}{2}}R_sR^{\frac{1}{2}}=U_s\Lambda U^H_s$. }

%$\frac{\textbf{{C}}^H\textbf{{C}}}{M}\xrightarrow{a.s.}{{R}}$ as $M \rightarrow \infty$.  $\frac{\textbf{{S}}^H\textbf{{S}}}{n}\xrightarrow{a.s.}{{R_s}}$.  $U_s$ is orthogonal matrix, they are results of eigenvalue decomposition, i.e. $R^{\frac{1}{2}}R_sR^{\frac{1}{2}}=U_s\Lambda U^H_s$, $U^H_{\textbf{{Y}}}U_{\textbf{{Y}}}\xrightarrow{a.s.}{\textbf{I}}_M$
\begin{proof}Please refer to Appendix B for detailed proof.
\end{proof}

%When the  vectors of $\{{\textbf{h}}_1,{\textbf{h}}_2,\cdots,{\textbf{h}}_{N}, {\textbf{g}}_1,{\textbf{g}}_2,\cdots,{\textbf{g}}_{N} \}$ are pairwisely independent with each other, we have diagnonal matrix ${\textbf{R}}$ such that $\frac{\textbf{{C}}^H\textbf{{C}}}{M}\xrightarrow{a.s.}{\textbf{R}}$ as $M \rightarrow \infty$. In addition, for arbitrary fixed ${\textbf{R}}_s=\frac{\textbf{{S}}^H\textbf{{S}}}{n}$, ${\textbf{R}}_{\textbf{{Y}}}=\frac{1}{M}P\{\textbf{{C}}{\textbf{R}}_s\textbf{{C}}^H  \}+\frac{\sigma^{2}}{M} \textbf{I}_M$ can be decomposed into
%\begin{equation}
%{\textbf{R}}_{\textbf{{Y}}}\xrightarrow{a.s.}U_{\textbf{{Y}}}diag\{\sigma^{2} \textbf{I}_{M-N-N}, \Lambda + \sigma^{2}\textbf{I}_{N+N} \}U^H_{\textbf{{Y}}}
%\end{equation} as $M \rightarrow \infty$, where $U_{\textbf{{Y}}}=\frac{1}{\sqrt{M}} [U_W, \textbf{{C}}{\textbf{R}}^{-\frac{1}{2}}U_d]$. $U_W  \in\mathbb{C}^{M\times (M-N-N)}$ has orthogonal columns, and spans null space of $[\textbf{{C}}{\textbf{R}}^{-\frac{1}{2}}U_d]$. $\Lambda$ is diagonal matrix and $U_d$ is orthogonal matrix, they are results of eigenvalue decomposition, i.e., $P{\textbf{R}}^{\frac{1}{2}}{\textbf{R}}_s{\textbf{R}}^{\frac{1}{2}}=U_d \Lambda U_d ^H.  U_{\textbf{{Y}}}^HU_{\textbf{{Y}}}\xrightarrow{a.s.} \textbf{I}_M$, as $M \rightarrow \infty$.
\end{mypro3}
%As long as $n$ is sufficient large, $\emph{\textbf{R}}_{\emph{\textbf{{Y}}}}$ could be approached by $\frac{1}{Mn}\emph{\textbf{{Y}}}\emph{\textbf{{Y}}}^H$ in the large $M$ limit.

\emph{Remark 2:}
Notice that $U_{\emph{\textbf{{Y}}}}$ is determined by $\emph{\textbf{{C}}}{{R}}^{-\frac{1}{2}}$ and $U_s$. $U_s$ hinges on the degree of correlation among rows of  $\textbf{\emph{S}}$. In  \cite{evd2}\cite{doubletrain}, all data streams are independent, i.e. , $U_s=\emph{\textbf{I}}$. Then $U_{\emph{\textbf{{Y}}}}$ is irrelevant to interference data. $U_{\emph{\textbf{{Y}}}}$ is thus used to directly eliminate interference from MUs. However, in this paper, due to  \emph{correlative} attacks, $U_s\neq\emph{\textbf{I}}$. The null space of MUs cannot be found from $U_{\emph{\textbf{{Y}}}}$, but the subspaces corresponding to MUs and LUs are united by $U_{\emph{\textbf{{Y}}}}$. This indicates that the MUs can directly manipulate $U_{\emph{\textbf{{Y}}}}$   by conducting a \emph{correlative} attack, hence past  work no longer applies  \cite{evd2}\cite{doubletrain}.

%\newtheorem{remark}{Remark}
%\begin{remark}
%Notice that $U_{\textbf{{Y}}}$ is determined by not only $\textbf{{C}}\emph{\textbf{R}}^{-\frac{1}{2}}$ but also $U_d$. $U_d$ hinges on the degree of correlation among rows of  $\textbf{\emph{S}}$. In \cite{wu}
%\end{remark}

As shown in (\ref{matrixform}), the received signals  $\textbf{\emph{Y}}$
are   mixtures of $\textbf{\emph{S}}$, where the mixing matrix $\textbf{\emph{C}}$ includes all channel vectors. Every element of noise distortion $\textbf{\emph{N}}$ follows $\mathcal{CN}(0, \sigma^{2})$, $\textbf{\emph{N}}\in\mathbb{C}^{M\times n}$.
This observation motivates us to achieve channel estimation by blind signal separation (BSS) approaches.
Nevertheless, due to the  attack, there is a correlation between $\emph{\textbf{a}}_j$ and $\emph{\textbf{b}}_{j_w}$, and the BS does not know  the statistical characteristics  of $\emph{\textbf{a}}_j$ and $\emph{\textbf{b}}_{j_w}$.
%{\color{blue}{$\left\{ a_j(t)\right\}$ and $\left\{ b_k(t)\right\}$ may be dependence, and the stochastic distribution of $\left\{ b_k(t)\right\}$}}
Traditional BSS techniques\cite{bca}\cite{volum} do not apply  to the attack scenario  considered in this paper.
We next propose a BSS technique that works well under  {\color{black}\emph{correlative} }attacks.
}

\begin{figure}
\centering
\includegraphics[width=0.50\textwidth]{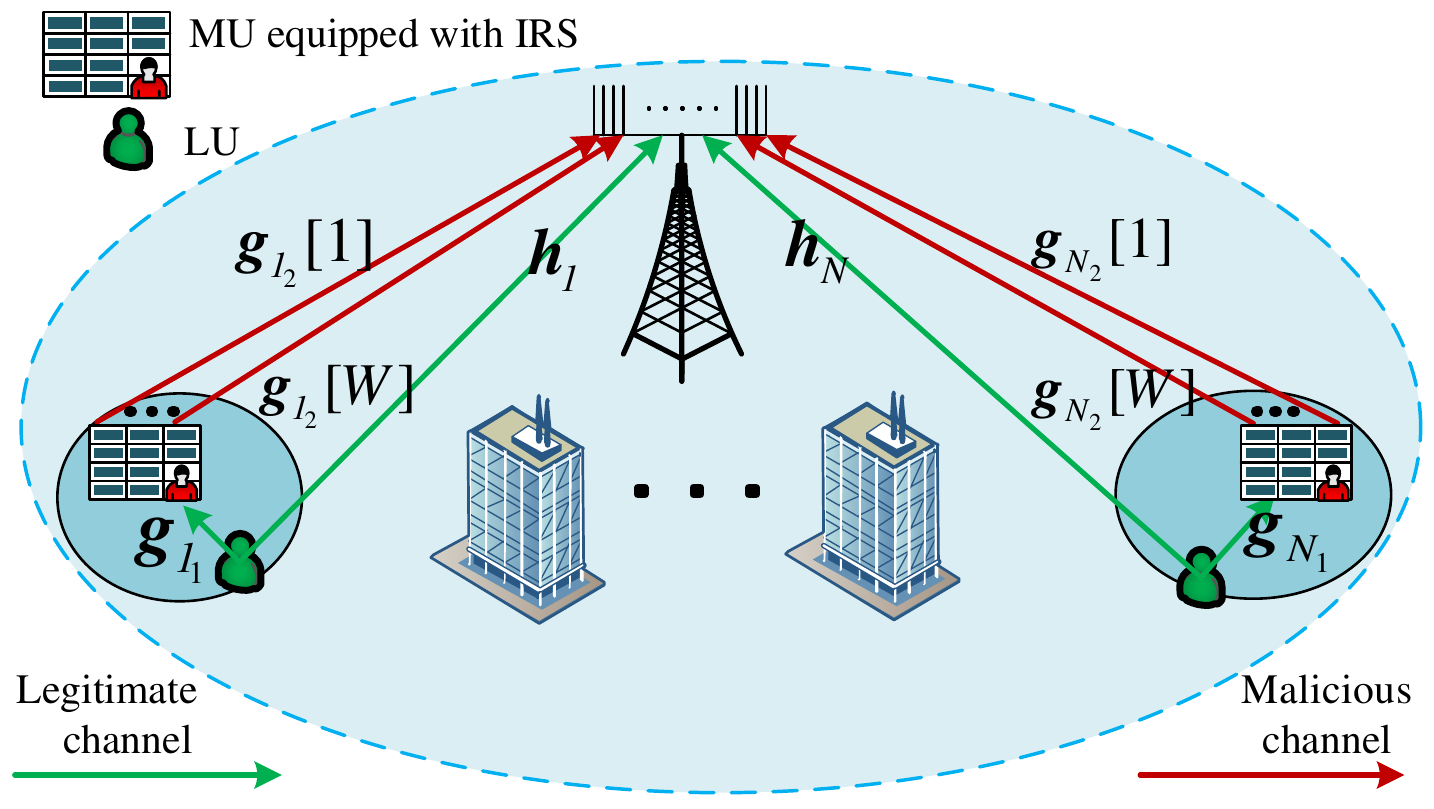}
\caption{{\color{black}System model including one BS {\color{black}equipped with $M$ antennas} and   $N$ group users. Each MU is equipped
with an IRS that includes $W$ elements.  Through the pilot and data phases, the MUs reflect  $W$ stream  pilot sequences and interference sequences to the BS.}}
\label{fig:attackmodel}
\end{figure}

\section{Signal Extraction and Channel Estimation}\label{Sep}
{\color{black}{
For a  \emph{correlative}  attack, we consider a geometric argument that is insensitive to  correlation.    For instance, the convex hull of ${\emph{\textbf{u}}}\textbf{\emph{C}} \textbf{\emph{S}}$ \cite{bca},
where ${\emph{\textbf{u}}}$ is the normalized vector  combination vector of $\textbf{\emph{C}} \textbf{\emph{S}} $, only depends on the alphabet
of ${\emph{\textbf{u}}}\textbf{\emph{C}} \textbf{\emph{S}} $, regardless of the correlation of $\textbf{\emph{S}} $.
%Let us use $\mathcal{L}\left(\cdot\right)$ denote the length of convex hull of its input sequence, namely convex perimeter.
 $\mathcal{L}\left({\emph{\textbf{u}}}\textbf{\emph{C}} \textbf{\emph{S}} \right)$ achieves its minimum  when ${\emph{\textbf{u}}}\textbf{\emph{C}} \textbf{\emph{S}} $ includes only one data stream rather than the mixture of several streams, where $\mathcal{L}\left(\cdot\right)$ denotes the length of the convex hull of its input sequence, i.e., the convex perimeter \cite{bca}.
Hence, the convex perimeter can be used for signal  extraction, which is also the basis of channel estimation.
However,  the BS receives only the noisy observation of $\boldsymbol{CS}$, i.e., $\textbf{\emph{Y}}$, rather than $\boldsymbol{CS}$ itself.
The convex perimeter is very sensitive to noise. As seen in Fig. \ref{fig:Qpskconv1}, the noise significantly changes the convex hull; hence, it
impacts the convex perimeter.   Proposition 2 provides an extractor  capable of distilling alphabets from  noisy observations.
%We first give Proposition 1 which indicates an extractor with capability of distilling eometric characteristics of a sequence from its noisy observation.

%Due to the existence of  \emph{correlative}  attack, we consider to utilize the DSs' geometric characteristic. More precisely, let us use $\mathcal{L}\left(\cdot\right)$ denote the length of convex hull of its input sequence, namely convex perimeter.  For instance, the convex perimeter of linear combined of DS remains unchanged regardless of correlation coefficients of $a_k$ and $b_k$.
%However, the BS cannot get the DS directly, but the superposition of the DS and noise is observed.
%Importantly, the convex perimeter is sensitive to noise. As illustrated in ~fig. \ref{fig:Qpskconv1}, the noise changes the convex hull significantly, and thus
%impacts the convex perimeter.
%The convex perimeter achieve its minima when includes only one symbol stream rather than the mixture of several streams  \cite{bca}.
%It indicates the convex perimeter could be used for symbol extraction, which is also the basis of channel estimation. Using the minima convex perimeter of the DS and noise to extract signal and then estimate channel \cite{bca}, but the estimation result under the noise influence. Thus, we first give Proposition 1 which indicates an extractor with capability of distilling alphabets from its noisy observation.
}}
\begin{figure}[h]
\centering
\includegraphics[width=0.5\textwidth]{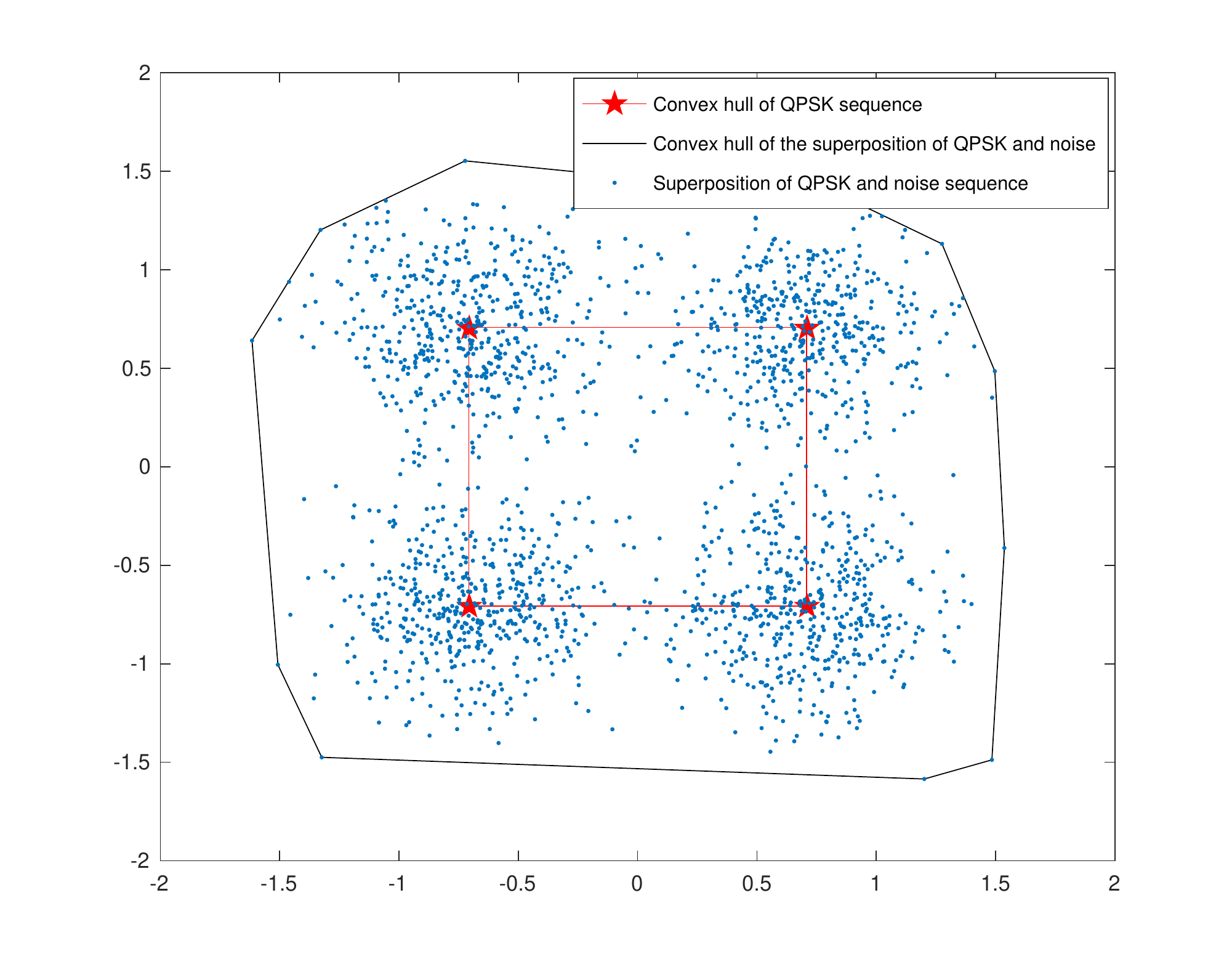}
\caption{Illustration of the   constellation of QPSK sequences (red dots) and QPSK and noise sequences (blue dots). Superimposed are the boundary and vertex  of its convex hull; the noise changes the convex hull significantly.}
\label{fig:Qpskconv1}
\end{figure}
%In order to estimate the channel, we first extract the signal. Under the attack and noise, we don' t know the stochastic propoerty of MUs and symbols of LUs and MUs are dependence. Fortunately, we find that the signal extraction correspond with the minimization of convex perimeter of TS, and the geometric characteristics of TS is sensitive to noise, which causes the performance of signal extraction deteriorate. We propose an extractor $\mathbb{{F}}$ that can get the true alphabets (TA) from noisy signal, TA is the Cartesian product of channel's alphabets  and signal's alphabets.
 \thispagestyle{empty}
{\color{black}{\newtheorem *{mypro2}{Proposition 2}
\begin{mypro2}
Let us denote the alphabet of a discrete and $n$-length i.i.d. sequence $V^n$  as $\mathcal{V}$. Another noise sequence $W^n$
 is independent of $V^n$. Then, from   $V^n+W^n$, there exists an extractor $\mathbb{{F}}$, by whose use, i.e., $\mathbb{{F}}(V^n+W^n)$,
$\mathcal{V}$ can be extracted in probability as $n$ approaches   infinity.
\end{mypro2}
\begin{proof}
Proposition 2 is proved by the  proposal of extractor $\mathbb{{F}}$ in  Appendix C.
%Due to space limitation, please refer to [Appendix C, \cite{zheng2020uplink}] for detailed proof of Proposition 2.

\end{proof}}}
We use the extractor $\mathbb{{F}}$ to distill $\mathcal{V}$ from $V^n+W^n$. Since $V^n$ is a discrete sequence, the convex perimeter of $\mathcal{V}$ is equivalent to that of $V^n$. We thus have  Corollary 1, which is based on Proposition 2.
{\color{black}
 \newtheorem{corollary}{Corollary}
\begin{corollary}
For a discrete and $n$-length i.i.d. sequence $V^n$,  there is another noise sequence $W^n$, which is independent of $V^n$. {\footnotesize $\mathcal{L}\{\mathbb{F}\{V^n+W^n\}\} \rightarrow \mathcal{L}\{V^n\}$} in probability as n approaches   infinity.
\end{corollary}
}
 {\color{black}{We next use $\mathbb{{F}}$  and $\mathcal{L}$ to extract signals and estimate channels.}}
%The observation signal of BS is the TS and noise. Next, we use the geometric characteristics to achieve the signal extraction and channel estimation. And we use  $\mathbb{{F}}$ to get the TA, then the alphabets are used in the signal extraction and channel estimation.   %We get the function of $\mathbb{{F}}$ in section \ref{noiseopt}.
%The signal that BS receives  in data transmission phase can be decomposed to as follows and the objective of the section is separate all $S_i$, and estimate all $C_i$.
%\begin{footnotesize}
%\begin{equation}
%\begin{aligned}
%\textbf{\emph{Y}}=\sum_{i=1}^{N_L+N_M}C_iS_i  +\textbf{\emph{N}}
%\end{aligned}
%\end{equation}%
%\end{footnotesize}

 \subsection{Signal Extraction}
 {\color{black}{Revisiting (\ref{matrixform}), $\textbf{\emph{Y}} $ is the superposition of $\textbf{\emph{C}} \textbf{\emph{S}}$
 and {\color{black}}  $\textbf{\emph{N}}$. Notice that signal extraction corresponds to the minimization of the convex perimeter of ${\emph{\textbf{u}}}\textbf{\emph{C}} \textbf{\emph{S}}$ \cite{bca}. Relying on our proposed extractor $\mathbb{{F}}$ to
 achieve the convex perimeter, we establish an optimization problem subject to the signal extraction vector, where $\emph{\textbf{u}} \in \mathbb{C}^{1\times M}$, }}
 {\color{black}{ \begin{equation}
\begin{aligned} \footnotesize \label{opt1}
\left [\hat{\emph{\textbf{u}}}\right]&=arg \left\{ \min \mathcal{L} \left\{   \mathbb{F} \left\{\emph{\textbf{u}}\textbf{\emph{Y}} \right\} \right\} \right\}\\
\quad \quad \quad &s.t.  \left\|\textbf{\emph{u}}\right\|_2=1.
\end{aligned}
\end{equation}}}
%{\color{black}{where ${\emph{\textbf{u}}}$ denotes the signal extraction vector, $\emph{\textbf{u}} \in \mathbb{C}^{1\times M}$
Since {$\emph{\textbf{u}}\textbf{\emph{Y}}=\emph{\textbf{u}}\textbf{\emph{C}} \textbf{\emph{S}}+\emph{\textbf{u}}\textbf{\emph{N}},$} {\color{black}according to Corollary 1,
{$\mathcal{L} \left\{   \mathbb{F} \left\{\emph{\textbf{u}}\textbf{\emph{Y}} \right\} \right\}\rightarrow\mathcal{L}\{\emph{\textbf{u}}\textbf{\emph{C}} \textbf{\emph{S}}\}$} in probability as $n$ approaches   infinity.
 The signal  extraction vector is achieved by minimizing {\footnotesize$\mathcal{L} \left\{\mathbb{F} \left\{\emph{\textbf{u}}\textbf{\emph{Y}} \right\} \right\}$}}} \cite{bca}. {\color{black}{As the contrast function of (\ref{opt1}) reduces the impact of noise, problem (\ref{opt1}) can be solved by  traditional  gradient descent    \cite{bca}.
Details can be found in Algorithm 1.

{\color{black}{The key difference of our work is the employment of $\mathbb{F}$
to reduce the impact of noise on the calculation of the convex perimeter.
Previous work investigated the noiseless scenario, obtaining a signal  extraction vector   by minimizing the convex perimeter of   $\emph{\textbf{u}}\textbf{\emph{Y}}$ \cite{bca}.
In   our model, due to the existence of noise, we propose $\mathbb{F}$ to obtain the convex perimeter of $\emph{\textbf{u}}\textbf{\emph{C}} \textbf{\emph{S}}$.
Simulations confirm that  $\mathbb{F}$ significantly enhances  extraction performance  in the presence of a correlation attack and noise.}}

{\color{black}{Based on (\ref{opt1}), the signal of one user %(LU or MU) \
{\color{black}is} extracted as }}
\begin{equation}  \label{signalextraction}
\emph{\textbf{s}}=\hat{\emph{\textbf{u}}}\textbf{\emph{Y}},
\end{equation}
where $ \emph{\textbf{s}} \in \mathbb{C}^{1\times n}$.
{\color{black}{We next {\color{black}estimate} one channel corresponding to the extracted $\emph{\textbf{s}}$.}}
{\color{black}{ \subsection{Channel Estimation}
%In this subsection, we focus on how to estimate the channel that correspond to the signal.
Without loss of generality, we let $ \emph{\textbf{c}} \in \mathbb{C}^{M\times1}$ denote the channel corresponding to the extracted $\emph{\textbf{s}}$.
For $m=1\cdots M$, we let $[\cdot]_m$   denote the  $m$th row of its input matrix or vector,
and rewrite $ \textbf{\emph{Y}}$ and $\textbf{\emph{R}}$ is the remainder signal.
 \begin{equation}  \label{dimin}
 [\textbf{\emph{Y}}]_m=[\textbf{\emph{R}}]_m+[\emph{\textbf{c}} ]_m\emph{\textbf{s}}.
 \end{equation}

Both $[\textbf{\emph{R}}]_m$ and $[\emph{\textbf{c}} ]_m\emph{\textbf{s}} $ are noisy observations that {\color{black}include} noise and discrete sequences. Thus,  $[\textbf{\emph{Y}}]_m$
is the noisy mixture of  sequences corresponding to $[\textbf{\emph{R}}]_m$ and $[\emph{\textbf{c}} ]_m\emph{\textbf{s}}$.
  $\mathcal{L}$ achieves its  local minimum value when its input is the alphabet of a single signal
rather than any mixture. Therefore, relying on $\mathbb{F}$,  $\emph{\textbf{c}}$ can be estimated by
% $S$ is  the signal that has extracted. $C$ is  the channel that we want to estimate,  $ C \in \mathbb{C}^{M\times1}$. In fact, this problem can be transformed  to a convex optimization problem, $m=1\cdots M$.
\begin{equation} \small \label{opt2}
[\hat{\emph{\textbf{c}}}]_m=arg \left\{ \min \mathcal{L} \left\{  \mathbb{F}\left ([\textbf{\emph{Y}}]_m-c'\emph{\textbf{s}}\right) \right\} \right\},  c' \in \mathbb{C}.
\end{equation}
To solve this problem, we also prove that the solution is in a finite and discrete set, which leads to an optimum solution when searching the finite and discrete set.
 \thispagestyle{empty}
\newtheorem*{mypro1}{Proposition 3}
\begin{mypro1}
The optimum solution to (\ref{opt2}) is included in a finite and discrete set,
$ \footnotesize{ \mathrm{Q_{m}}=\left\{ q\:|\: q=\frac{\mathsf{y}-\mathsf{y}'}{\mathsf{z}-\mathsf{z}'},\mathsf{z}\neq\mathsf{z}',\mathsf{y}\neq\mathsf{y}',\right.}$ $\left.\mathsf{y},\mathsf{y}'\in \mathcal{Y}_m ,\mathsf{z},\mathsf{z}'\in \mathcal{Z} \right\} $, where  \footnotesize $\mathcal{Y}_m= \mathbb{F}\left\{ [\textbf{{Y}}]_m\right\}$,$\mathcal{Z}=\mathbb{F}\left\{{\textbf{s}}\right\}$.
\end{mypro1}
\begin{proof}
Please refer to Appendix D for detailed proof.
\end{proof}

The key feature of (\ref{opt2}) is the use of our proposed extractor $\mathbb{F}$  in the contrast function of (\ref{opt2}), and in Proposition 3 to locate a solution.
Previous work \cite{letter} only considers a noiseless scenario and implements no denoising measures.

Based on Proposition 3, we can estimate $\textbf{\emph{c}}$ as $\hat{\textbf{\emph{c}}}$ from (\ref{opt2}).
Then, with $\hat{\textbf{\emph{c}}}$ and the extracted $\textbf{\emph{c}}$, the contribution of $\textbf{\emph{s}}$ can be removed from $\textbf{\emph{Y}}$.
After the deduction of $\hat{\textbf{\emph{c}}}\textbf{\emph{s}}$ from $\textbf{\emph{Y}}$, let us repeat the signal extraction and channel estimation, as presented in {Algorithm 1},
until all channels are estimated.

In Algorithm 1, steps 3$\sim$12 solve the optimization problem  (\ref{opt1})  by   gradient descent. The resulting vector  $\hat{\emph{\textbf{u}}}$ is  used for signal extraction in step 13. In steps 14$\sim$19, optimization problem ({\ref{opt2}}) is solved by searching the discrete solution set given by Proposition 3. After   $\textbf{\emph{s}}$ and $\hat{\textbf{\emph{c}}}$ are obtained, we deduct $\hat{\textbf{\emph{c}}}\textbf{\emph{s}}$ from $\textbf{\emph{Y}}$, and iteratively run channel estimation.
%the step 3$\sim$13 correspond to getting the signal extraction vector $\hat{\emph{\textbf{u}}}$ (\ref{opt1}), and the step 8$\sim$12 correspond to the traditional gradient descent method,  the step 16$\sim$20 correspond to ({\ref{opt2}}) that estimate channels.

\begin{algorithm}
{
\caption{  Signal  Extraction and Channel  Estimation}%算法名字
\% $i$:  the $i$th signal extraction and channel estimation\; %\;用于换行
\% $\mathbb{F}$: our proposed extractor given by  {Algorithm 2} in [Appendix C, \cite{zheng2020uplink}]\;
\%  $convhull$: the function of getting  the convex points of its input sequence\;
\% $\mathbb{P}$: the function that  finds $ \emph{\textbf{u}}\textbf{\emph{Y}}(p_1)$, $ \emph{\textbf{u}}\textbf{\emph{Y}}(p_2)$,  $ \emph{\textbf{u}}\textbf{\emph{Y}}(\cdots)$, and $ \emph{\textbf{u}}\textbf{\emph{Y}}(p_{n'})$ nearest to $\textbf{\emph{y}}(k_1)$, $\textbf{\emph{y}}(k_2)$, $\textbf{\emph{y}}(\cdots)$, and $\textbf{\emph{y}}(k_{n'})$, respectively\;
\LinesNumbered %要求显示行号
{\color{black}\KwIn{$\textbf{\emph{Y}}$, $N$, $W$, $M$}}%输入参数
\KwOut{signals $\textbf{\emph{s}}_i$ and channels $\hat{\textbf{\emph{c}}}_i$, for $i=1,2,\cdots, N_L+N_M$}%输出

\For{{$i=1:1:N+WN$}}{
Initialization: $\emph{\textbf{u}}(1)=1$, $\emph{\textbf{u}}(2:M)=0$\;
\For{{iter=1:1:until $\mathcal{L} \left\{   \mathbb{F} \left\{\emph{\textbf{u}}\textbf{\emph{Y}} \right\} \right\}$ minimum}}{
  $\textbf{\emph{y}}= \mathbb{F}(\emph{\textbf{u}}\textbf{\emph{Y}})$  \% $\textbf{\emph{y}}$ is the alphabet of $\emph{\textbf{u}}\textbf{\emph{Y}}$\

$[k_1,k_2,\cdots,k_{n'}]= convhull(\Re\left\{\textbf{\emph{y}} \right\},\Im\left\{\textbf{\emph{y}} \right\})$\

$\mathcal{L}(\textbf{\emph{y}})=\sum_{i=2}^{n'} \left\| \textbf{\emph{y}}(k_i)-\textbf{\emph{y}}(k_{i-1})\right\|_2 $\

$ [p_1,p_2,\cdots,p_{n'}]=\mathbb{P}\left\{[k_1,k_2,\cdots,k_{n'}]\right\}$ \

$ \emph{\textbf{Wp}}=\sum_{i=2}^{n'}{ \left\{\textbf{\emph{Y}}\left(:,p_i\right) -\textbf{\emph{Y}}\left(:,p_{i-1}\right)\right\}}$ \\ $\quad\quad\quad{\left\{\textbf{\emph{Y}}\left(:,p_i\right) -\textbf{\emph{Y}}\left(:,p_{i-1}\right)\right\}^{H}}/(\left\| \textbf{\emph{y}}(k_i)-\textbf{\emph{y}}(k_{i-1})\right\|_2)$\

$\emph{\textbf{g}}=({\frac{1}{2}\emph{\textbf{Wp}}{\emph{\textbf{u}}}-\emph{\textbf{u}} \mathcal{L}(\textbf{\emph{y}})})/{ \left\|\emph{{\textbf{u}}}\right\|_2}$ \

$\mu= 1/({2{\left\|  \emph{\textbf{g}}\right\|^{2}_2}}$) \

$\emph{\textbf{u}}=({\emph{\textbf{u}}-\mu  \emph{\textbf{g}}})/{{ \left\|\emph{{\textbf{u}}}-\mu  \emph{\textbf{g}}\right\|_2}}$ \

}
%　　Initialization: $\emph{\textbf{u}}(1)=1$, $\emph{\textbf{u}}(2:M)=0$\;
%　　\For{{iter=1:1:until $\mathcal{L} \left\{   \mathbb{F} \left\{\emph{\textbf{u}}\textbf{\emph{Y}} \right\} \right\}$ minimum}}{
  %       $\textbf{\emph{y}}= \mathbb{F}(\emph{\textbf{u}}\textbf{\emph{Y}})$  \% $\textbf{\emph{y}}$ is the alphabet of $\emph{\textbf{u}}\textbf{\emph{Y}}$\

%$[k_1,k_2,\cdots,k_{n'}]= convhull(\Re\left\{\textbf{\emph{y}} \right\},\Im\left\{\textbf{\emph{y}} \right\})$\

 %$\mathcal{L}(\textbf{\emph{y}})=\sum_{i=2}^{n'} \left\| \textbf{\emph{y}}(k_i)-\textbf{\emph{y}}(k_{i-1})\right\|_2 $\

%$ [p_1,p_2,\cdots,p_{n'}]=\mathbb{P}\left\{[k_1,k_2,\cdots,k_{n'}]\right\}$ \

%$ \emph{\textbf{Wp}}=\sum_{i=2}^{n'}{ \left\{\textbf{\emph{Y}}\left(:,p_i\right) -\textbf{\emph{Y}}\left(:,p_{i-1}\right)\right\}}$\\$ \quad  {\left\{\textbf{\emph{Y}}\left(:,p_i\right) -\textbf{\emph{Y}}\left(:,p_{i-1}\right)\right\}^{H}}/(\left\| \textbf{\emph{y}}(k_i)-\textbf{\emph{y}}(k_{i-1})\right\|_2)$
%\

%$\emph{\textbf{g}}=({\frac{1}{2}\emph{\textbf{Wp}}{\emph{\textbf{u}}}-\emph{\textbf{u}} \mathcal{L}(\textbf{\emph{y}})})/{ \left\|\emph{{\textbf{u}}}\right\|_2}$ \

%$\mu= 1/({2{\left\|  \emph{\textbf{g}}\right\|^{2}_2}}$) \

%$\emph{\textbf{u}}=({\emph{\textbf{u}}-\mu  \emph{\textbf{g}}})/{{ \left\|\emph{{\textbf{u}}}-\mu  \emph{\textbf{g}}\right\|_2}} $                                                                       \
 % }
$\textbf{\emph{s}}_i=\hat{\emph{\textbf{u}}}\textbf{\emph{Y}}$ (\ref{signalextraction}) \

$\mathcal{Z}_i= \mathbb{F} (\textbf{\emph{s}}_i)$\

\For {$m=1:1:M$}{
$ \mathcal{Y}_m= \mathbb{F} ([\emph{\textbf{Y}}]_m)$ \
	
$\mathrm{Q_{m}}= \big\{q\:|\: q=\frac{\mathsf{y}-\mathsf{y}'}{\mathsf{z}-\mathsf{z}'},\mathsf{z}\neq\mathsf{z}',\mathsf{y}\neq\mathsf{y}', \mathsf{y},\mathsf{y}'\in\mathcal{Y}_m,   \mathsf{z},\mathsf{z}'\in \mathcal{Z}_i \big\} $ \

$[\hat{\textbf{\emph{c}}}_i]_m=arg \left\{ \min \mathcal{L} \left\{ \mathbb{F}( [\textbf{\emph{Y}}]_m-c'\textbf{\emph{s}}_i)\right\} \right\} $, $c'\in  \mathrm{Q_{m}}$ (\ref{opt1}) \
}
$ \textbf{\emph{Y}}=\textbf{\emph{Y}}-\hat{\textbf{\emph{c}}}_i{\textbf{\emph{s}}_i}$

}}
\end{algorithm}

\subsection{Channel Identification}
{\color{black}{Note that the proposed signal extraction depends on the minimum of  $\mathcal{L} \left\{\mathbb{F} \left\{\emph{\textbf{u}}\textbf{\emph{Y}} \right\} \right\}$, where $\mathcal{L} \left\{\mathbb{F} \left\{\emph{\textbf{u}}\textbf{\emph{Y}} \right\} \right\}$ remains unchanged when the angles of its input are rotated.  Optimization problem (\ref{opt1})  just indicates that the extracted signal belongs to one user, but it cannot determine which user corresponds to the extracted signal. Hence, order ambiguity   exists.
%This property results phase ambiguity of  the signal extraction. still can reach minimum, then $\emph{\textbf{s}}=\hat{\emph{\textbf{u}}}\textbf{\emph{Y}}$, $\emph{\textbf{s}}$ may be phase ambiguity. And the extraction signal  belongs to one of the mixed signal, which is order ambiguity. The channel estimation depending on the signal extraction, so we could estimate all channels up to a phase and order ambiguity.
%After channel estimation, we could {\color{black}estimate} all channels up to a phase and order ambiguity.

Such ambiguities widely exist in BSS-based work \cite{bca} \cite{iden}. Previous work \cite{iden}  assumes that the phase and order ambiguities are resolved perfectly by outdated estimate results. We similarly assume perfect channel identification.

%In last subsection, we recover channel $[\emph{\textbf{H}}, \emph{\textbf{G}}]$ up to a phase and order ambiguity. To be more precise, the order of separated channels is unknown.  And the estimated channels $C=[\hat{\emph{\textbf{H}}}, \hat{\emph{\textbf{G}}}]$ could correspond to any of existing channels.  We use permutation matrix $\mathcal{Q}$ and phase rotation matrix to characterize the order and phase ambiguity.  Where $\mathcal{Q}=[q_1,\cdots,q_{N_L+N_m}]$, and $\mathcal{Q} \in \mathbb{C}^{{(N_L+N_M)}\times {(N_L+N_M)}}$.  For $ l=1,\cdots,N_L$, $q_l$ has  only one nonzero element of $1$. The ambiguity problem widely exists in a vast collection in BSS techniques \cite{iden}. And the problem is generally resolved by utilizing  additional information, for example, by using outdated estimate result. %And the identification is beyond scope of this paper.
%gi的身份hi地方

%\section{method of noise reduction}  \label{noiseopt}

\section{experimental results} \label{result}
%We consider i.i.d. Rayleigh fading channels with a $M\times2N$ channel matrix.
{\color{black}As the system model shows, there are $N$ group users, each group includes one LU and one MU, and each MU is equipped with an IRS with $W$ elements that can randomly reflect $W$ stream signal sequences. The channel is  i.i.d. Rayleigh fading, with an $M\times (N+WN)$ channel matrix.  {\color{black}Without loss of generality, we consider a massive MIMO system with $M=128$ antennas, and  attack scenarios of $N=2$,  $W=2$, $N=1$, $W=3$ and $N=1$, $W=1$, as shown in~Figs. \ref{fig:6user}, \ref{fig:4user}, and \ref{fig:2user}, respectively. The  independent symbols of LUs are drawn from a BPSK constellation,}
{\color{black}and we assume that the MUs conduct the attack  according to  (\ref{coree}). When  $W=1$, we set $\Pr\{\phi_{j_1}=0\}=0.8$,  and then  the correlation coefficient of $\emph{\textbf{a}}_j$ and $\emph{\textbf{b}}_{j_1}$ is $0.6$}. When $W=2$, we set $\Pr\{\phi_{j_1}=0\}=0.6$, $\Pr\{\phi_{j_2}=0\}=0.7$. Then the correlation coefficients of $\emph{\textbf{a}}_j$ and $\emph{\textbf{b}}_{j_1}$ and of $\emph{\textbf{a}}_j$ and $\emph{\textbf{b}}_{j_2}$ are $0.2$} and $0.4$}, respectively. When $W=3$, we set $\Pr\{\phi_{j_1}=0\}=0.6$, $\Pr\{\phi_{j_2}=0\}=0.7$, $\Pr\{\phi_{j_3}=0\}=0.8$. Then the correlation coefficients of $\emph{\textbf{a}}_j$ and $\emph{\textbf{b}}_{j_1}$, $\emph{\textbf{a}}_j$ and $\emph{\textbf{b}}_{j_2}$, and $\emph{\textbf{a}}_j$ and $\emph{\textbf{b}}_{j_3}$ are $0.2$},    $0.4$}, and    $0.6$}, respectively.  {\color{black}The BS estimates channels and achieves CSI. According {\color{black} to} the achieved CSI,  BS uses zero forcing (ZF) detection to get the signal-to-interference-and-noise ratio (SINR) of all users.\thispagestyle{empty} The normalized mean square error (NMSE) of the channel estimation and the bit error rate (BER) of the separation signal are also selected as   performance metrics.
We simulate and compare the performance of our proposed method and those based on bounded component analysis (BCA)  \cite{bca}
and EVD \cite{evd2}.} We also simulate the performance    achieved {\color{black} under }perfect CSI.} {\color{black} Since we consider multiple users,   to evaluate the performance of every user, we use the mean-SINR, min-SINR, max-NMSE, mean-NMSE, and min-NMSE, where ``mean'', ``min'', and ``max''  represent the average, worst, and best  performance metrics over all users.}
\thispagestyle{empty}

Because the EVD-based  method depends on the  transmission power gap between LUs and MUs   to get the separability of eigenspaces,  we set the path-losses of MUs  less {\color{black}than} LUs, that means the interference of MUs in our proposed method is much stronger than that of EVD-based method. We use  EVD-0.3 and EVD-0.5 denote the path-losses of MUs are 0.3 and 0.5, respectively.

\thispagestyle{empty}

\color{black}%In \cite{evd2}, better channel estimation performance depends on the separation of characteristic subspace and the data streams are independent, so the authors set the path loss of LUs and MUs are different. In this paper, we set the path loss of MUs are 0.3 and 0.5, respectively.
In~Figs. \ref{fig:6user}, \ref{fig:4user}, and \ref{fig:2user}, the path loss of MUs in the proposed method is $1$. %and the signals correlation coefficient of LUs and MUs is $0.6$.
Although the interference of  MUs in our proposed method was greater than the EVD-based method, the proposed method  performs better  than the EVD-based method, and our  method performs close to  the perfect CSI. Specifically, it is observed  in Figs. \ref{fig:6user}, \ref{fig:4user},  and \ref{fig:2user} that as the the correlation coefficient is fixed, the performance of the EVD-based method remains almost unchanged despite an increase in the signal-to-noise ratio (SNR). {\color{black}In contrast,  in Fig. \ref{fig:2usercor},  we consider $N=1, W=1$, and the performance of the EVD-based method  changes significantly when the correlation coefficient of  $\emph{\textbf{a}}_j$ and $\emph{\textbf{b}}_{j_1}$ increases from $0$ to $0.8$. Fig. \ref{fig:2usercor} presents the performance {\color{black} of} $N=1$ and  $W=1$ under varying correlation coefficients with an SNR of 16 dB. }The proposed method has better performance than the EVD-based method with different correlation coefficients.   This is consistent with  Proposition 1, indicating that in the presence of a \emph{correlative} attack, the signals of LUs and MUs  no longer  lie in distinct eigenspaces of the received signal matrix in the BS. Instead, the subspace of the attack signals overlaps with the eigenspace corresponding to the LUs, thus leading to attack leakage when the EVD-based method employs eigenvectors corresponding to the LUs for the received signal projection.

 In our proposed method, we consider reducing the impact of noise and use  geometric properties to overcome the impact of a \emph{correlative} attack. The performance increases as the SNR increases in Figs. \ref{fig:6user}, \ref{fig:4user}, and \ref{fig:2user}, and is unchanged in a certain range as the correlation coefficient increases in Fig. \ref{fig:2usercor}. Specifically, in  Fig. \ref{fig:2subfig:acoe}, it is observed that when the correlation coefficient is $0.6$,  the SINR of the proposed method is better than that of the EVD-based method by more than $5$ dB. %The performance of EVD-based increases with the decrease of correlation coefficient.
The EVD-based method has  better performance when the pass losses of MUs are less.  This indicates that the stronger the attack signals, the worse the performance is of the EVD-based  method. This  could be
because the EVD-based method attempts to eliminate attack signals as interference.  The proposed method treats attack signals as those of regular users, rather
than interference. We also estimate attack signals and channels instead of  eliminating them as interference. Thus the proposed method  outperforms the EVD-based method under much stronger attacks.
%The performance of EVD-based satisfies the Proposition 1.
%In \cite{evd2}, the authors don't consider the SER, so we omit it in our paper.}

{\color{black}Next, in~Figs. \ref{fig:6user}, \ref{fig:4user}, and \ref{fig:2user}, we  present the performance of the BCA method, and we see that the proposed method  performs much better. For instance, it is observed  in Fig. \ref{fig:6user} that the mean-SINR and min-SINR of the proposed method are better than those of the BCA method by more than $5$ dB.  The NMSE of the proposed method is better than that of the BCA method, especially at low SNRs. %The SER still has better performance than BCA.
 }
 \thispagestyle{empty}
%\begin{figure}
%\subfigure[]{
%\label{fig:6subfig:a}
%\includegraphics[width=0.28\textwidth]{6usera}}
%\subfigure[]{
%\label{fig:6subfig:b}
%\includegraphics[width=0.28\textwidth]{6userb}}
%\caption {(a) SINR  and (b) NMSE  of    proposed,  BCA, and  EVD-based methods versus   SNR with $N=2$,  $W=2$. The correlation between signal  sequences of LU and   MU are fixed at $0.2$ and $0.4$, respectively.}
%\label{fig:6user}
%\end{figure}

\begin{figure}[htbp]
\subfigure[]{
\label{fig:6subfig:a}
\begin{minipage}[t]{0.5\linewidth}
\includegraphics[height=1.2in,width=1.5in]{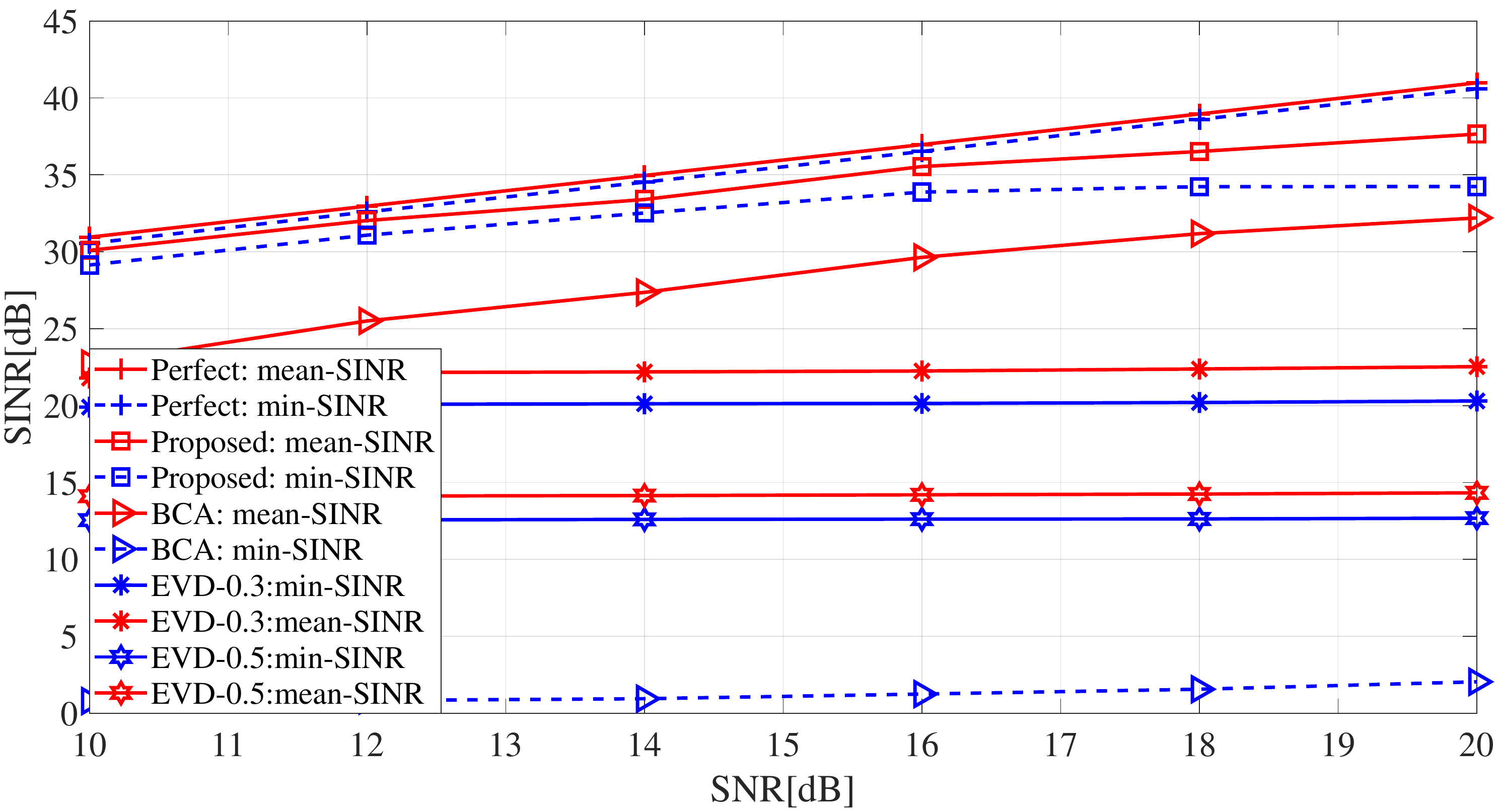}
\end{minipage}
}
\subfigure[]{
\label{fig:6subfig:b}
\begin{minipage}[t]{0.425\linewidth}
\includegraphics[height=1.2in,width=1.5in]{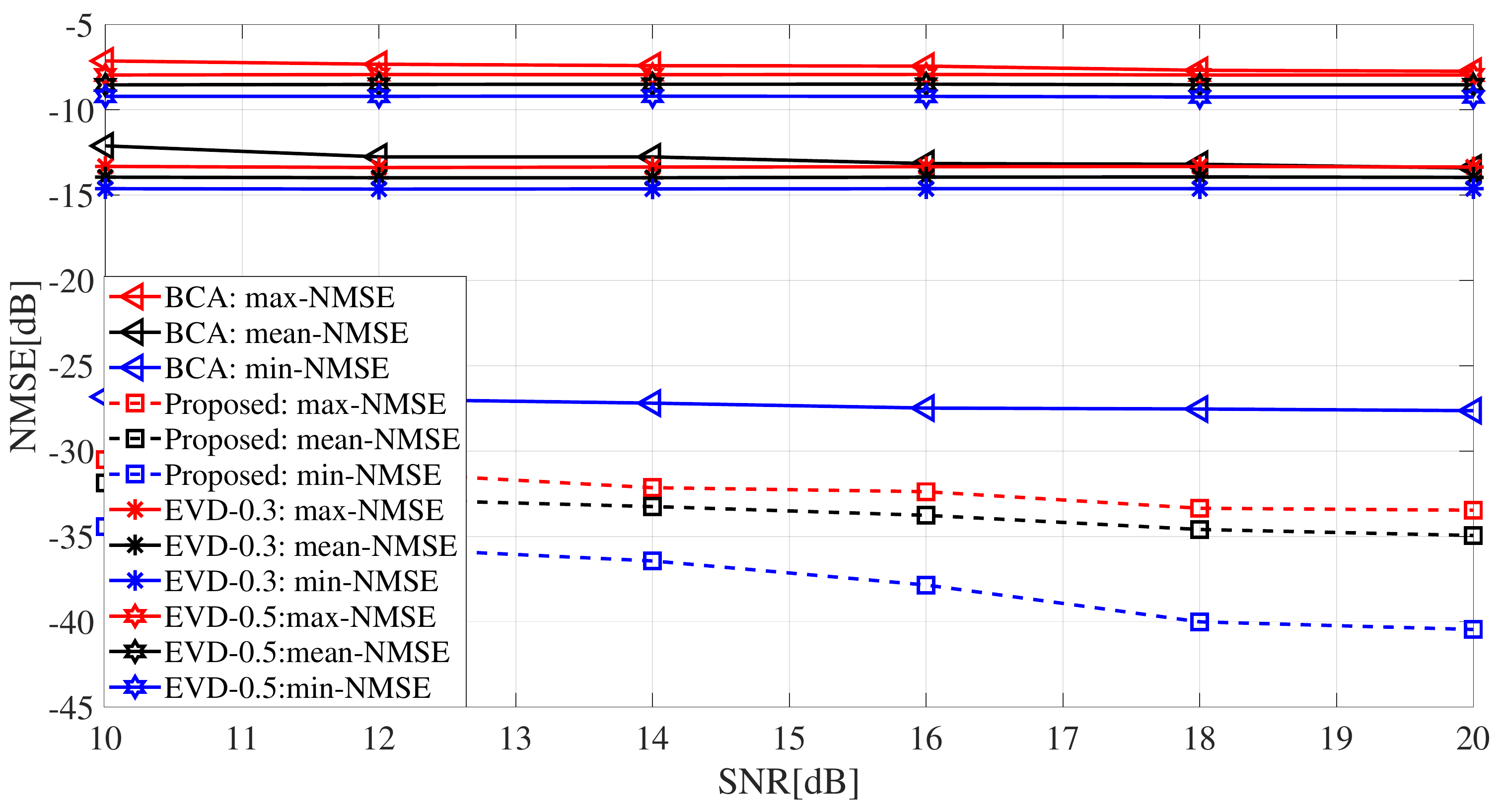}
\end{minipage}
}
\centering
\caption{ (a) SINR  and (b) NMSE  of    proposed,  BCA, and  EVD-based methods versus   SNR with $N=2$,  $W=2$. The correlation between signal  sequences of LU and   MU are fixed at $0.2$ and $0.4$, respectively.}
\label{fig:6user}
\end{figure}

\thispagestyle{empty}
%Next, we further consider  $4$ users, including $2$ MUs and $2$ LUs.  The performance  is shown in~Fig. \ref{fig:4user}.  We can find that the indexes of proposed method are better than BCA.

\begin{figure}[htbp]
\subfigure[]{
\label{fig:4subfig:a}
\begin{minipage}[t]{0.5\linewidth}
\includegraphics[height=1.2in,width=1.5in]{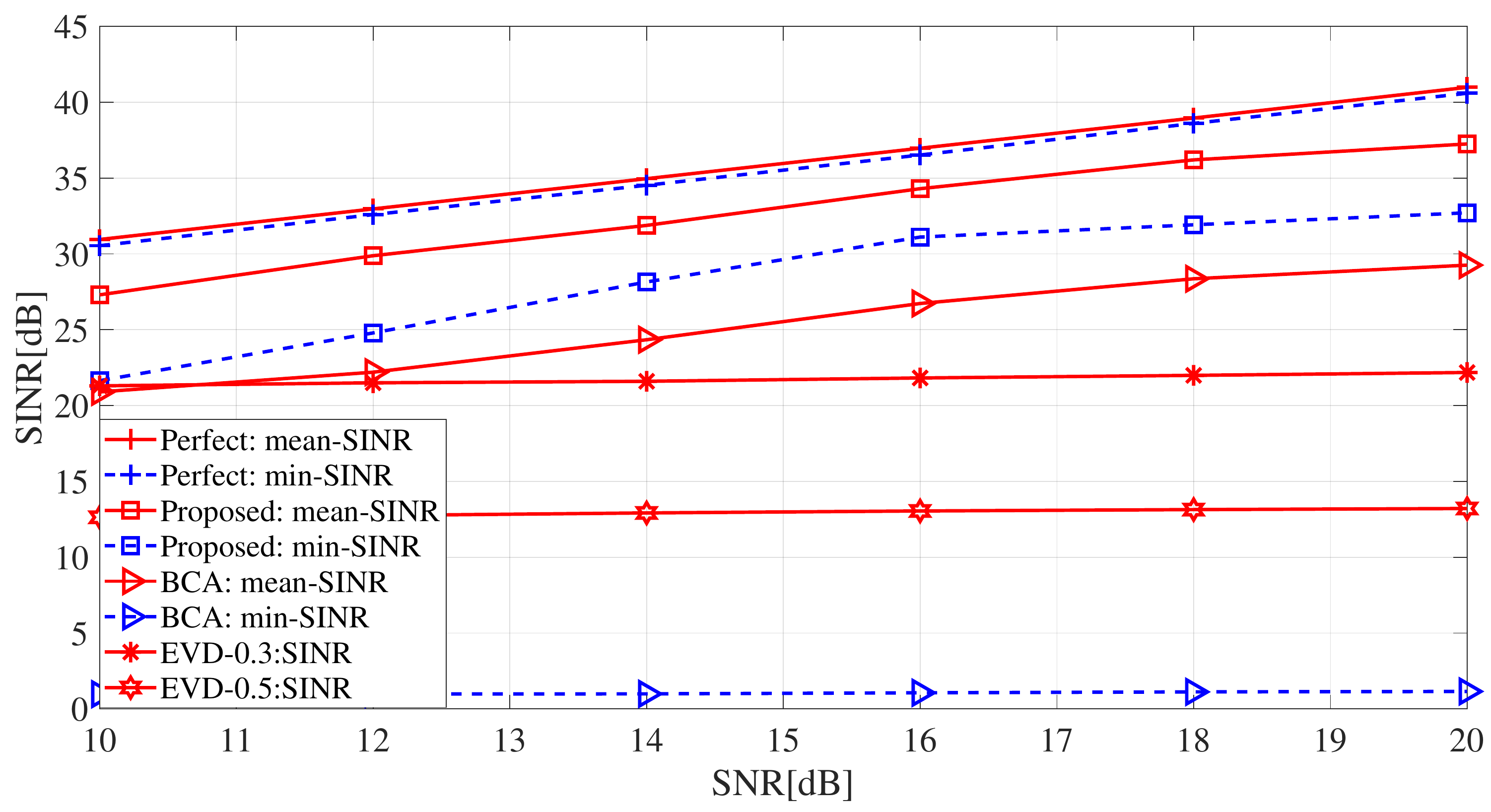}
\end{minipage}
}
\subfigure[]{
\label{fig:4subfig:b}
\begin{minipage}[t]{0.425\linewidth}
\includegraphics[height=1.2in,width=1.5in]{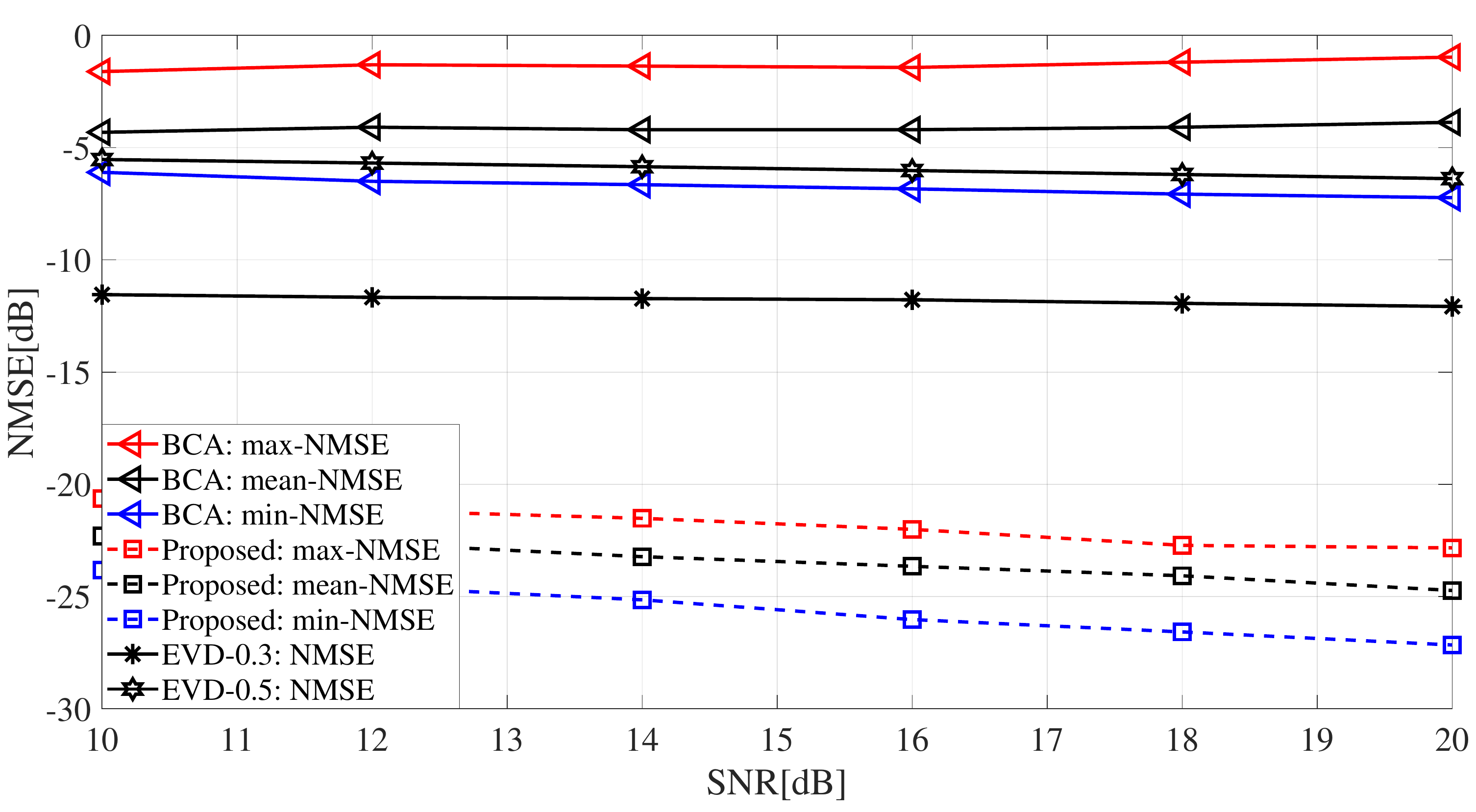}
\end{minipage}
}
\centering
\caption{(a) SINR  and (b) NMSE  of   proposed,  BCA, and  EVD-based methods versus  SNR with $N=1$,  $W=3$. The correlation between signal  sequences of LU and  MU are fixed at $0.6$, $0.7$ and $0.8$. }
\label{fig:4user}
\end{figure}

\thispagestyle{empty}
%Under the same experimental parameters, we further consider the system with  $1$ MU and $1$ LU, the performance in~Fig. \ref{fig:2user}.  Due to there are only  $2$ users, the interference between $2$ users is lower than  $6$ users, the gap with the perfect state decrease.
{\color{black}
To study the influence of signal correlation on  performance, in ~Fig. \ref{fig:2subfig:acoe},
%we simulate performance for two-user system including 1 LU and 1 MU under varying correlation coefficient in the SNR of 16.
it is observed that the mean-SINR and min-SINR of our proposed method outperform the BCA method by more than $5$ dB, and the NMSE of the  proposed method {\color{black}outperforms} that of the BCA method.  {\color{black} Fig. \ref{sersnr}  shows the BER performance of $N=2$,  $W=2$; $N=1$,  $W=3$; $N=1$,  $W=1$  for the proposed method and BCA method.  Fig. \ref{sercoe} shows the performance of $N=1$,  $W=1$ with different correlation coefficients.  The proposed method outperforms the BCA method in any case. }

   We further {\color{black}discover} that the performance of both methods,  especially the BCA method, will deteriorate as the correlation coefficient increases.  Actually, the BCA method works well in a noiseless scenario.
This indicates that the BCA method is sensitive to noise,  because it is based on geometric properties of desired signals.
The existence of noise changes the shape of the convex hull of  {\color{black}desired signals}.  Consequently,
geometric properties cannot be captured exactly in the presence of  noise.
Therefore, the existence of Gaussian noise damages
the performance of the BCA method  against a dependence attack.
In contrast, the performance of our proposed method changes  little as the correlation of users'  symbols increases.
Our   method considers the reduction of the impact of noise,
as mentioned above, thus  correlation does not significantly degrade its performance.
}

\thispagestyle{empty}

{\color{black} In summary, based on our simulation results, the proposed method  outperforms the BCA   and  {\color{black}EVD-based methods in the sense of SINR, NMSE, and BER}. }
%With the correction coefficient increase, the distill ability of extractor $\mathbb{{F}}$ weaken, causes the performance of proposed method deteriorate.  For BCA, when the correction coefficient increase, the noise influences the geometric characteristics of DS more seriously.  Then the performance of BCA deteriorates greatly.
%\begin{figure}[h]
%\centering
%\subfigure[]{
%\label{fig:2subfig:a}
%\includegraphics[width=0.28\textwidth]{2usera}}
%\subfigure[]{
%\label{fig:2subfig:b}
%\includegraphics[width=0.28\textwidth]{2userb}}
%\caption{(a) SINR  and (b) NMSE  of   proposed,  BCA, and  EVD-based methods versus  SNR with $N=1$, $W=1$. The correlation between LU and MU is fixed at $0.6$.}
%\label{fig:2user}
%\end{figure}

\begin{figure}[htbp]

\subfigure[]{
\label{fig:2subfig:a}
\begin{minipage}[t]{0.5\linewidth}
\includegraphics[height=1.2in,width=1.5in]{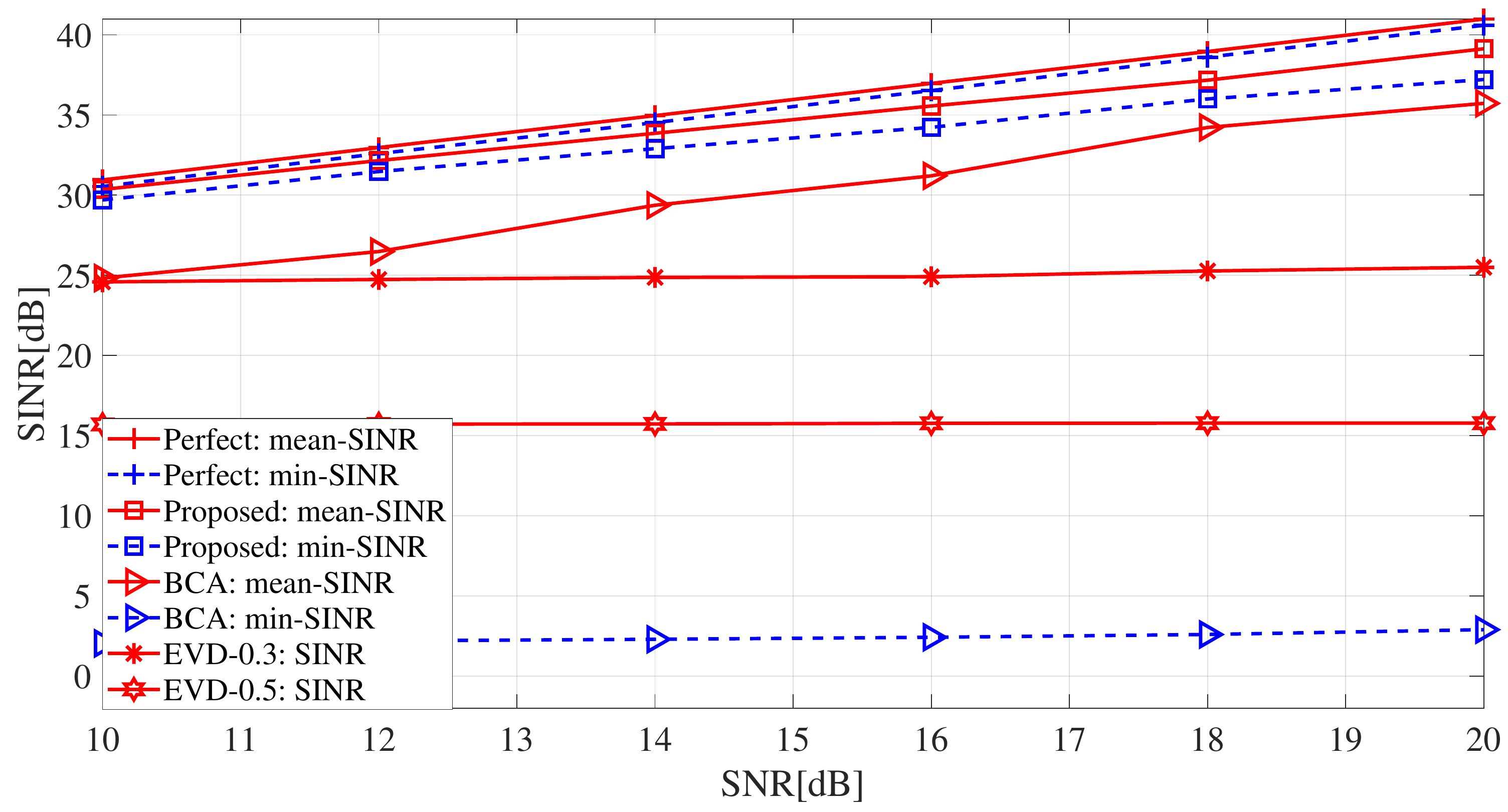}
\end{minipage}
}
\subfigure[]{
\label{fig:2subfig:b}
\begin{minipage}[t]{0.425\linewidth}

\includegraphics[height=1.2in,width=1.5in]{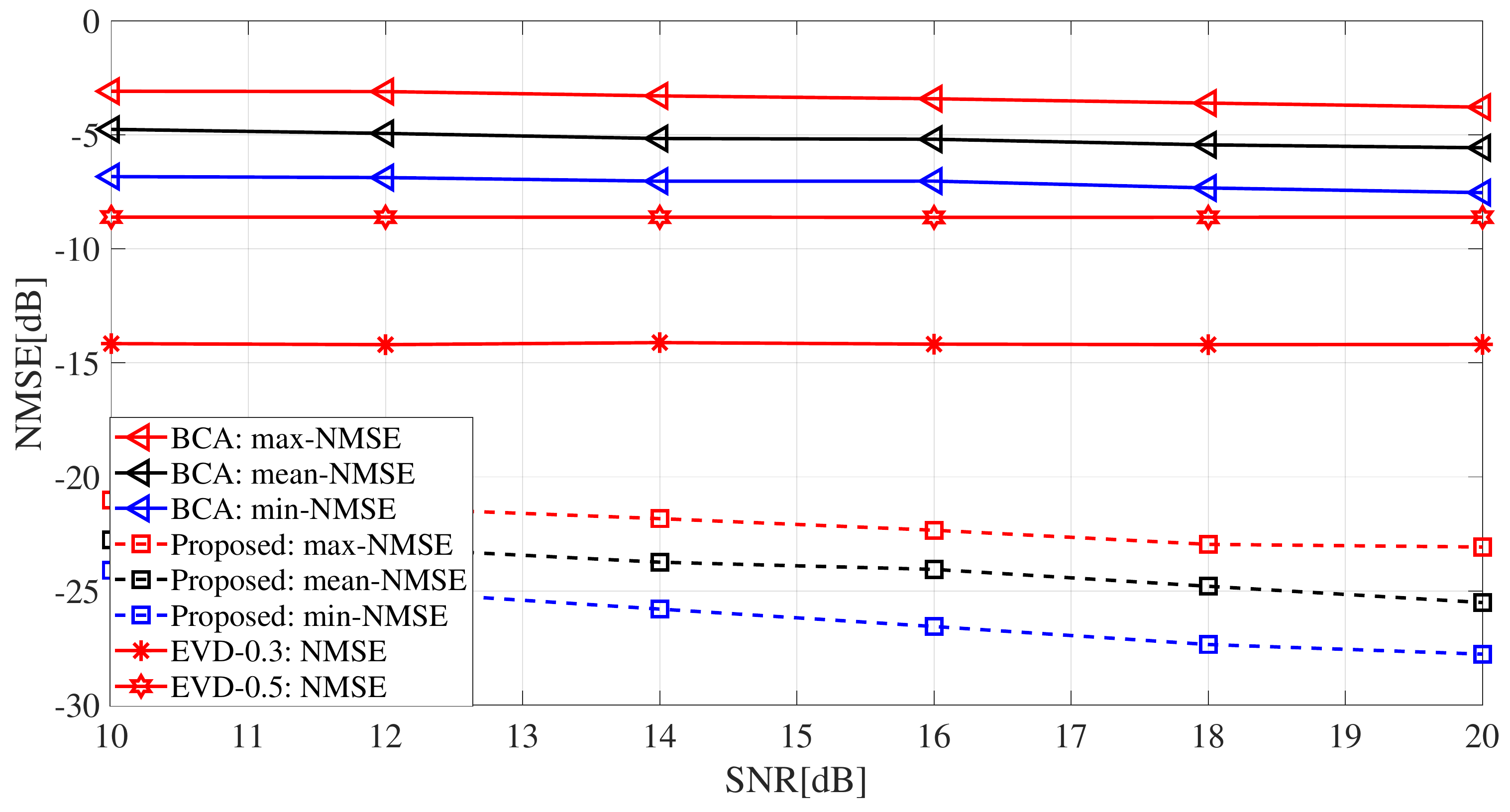}
\end{minipage}
}
\centering
\caption{ (a) SINR  and (b) NMSE  of   proposed,  BCA, and  EVD-based methods versus  SNR with $N=1$, $W=1$. The correlation between LU and MU is fixed at $0.6$.}
\label{fig:2user}
\end{figure}
\thispagestyle{empty}

%\begin{figure}[h]
%\centering
%\subfigure[]{
%\label{fig:2subfig:acoe}
%\includegraphics[width=0.28\textwidth]{2usercora}}
%\subfigure[]{
%\label{fig:2subfig:bcoe}
%\includegraphics[width=0.28\textwidth]{2usercorb}}
%\caption{(a) SINR  and (b)  NMSE  of   proposed,  BCA, and  EVD-based methods versus  correlation coefficient  with $N=1$, $W=1$. The SNR is fixed at 16 dB.}
%\label{fig:2usercor}
%\end{figure}
\begin{figure}[htbp]

\subfigure[]{
\label{fig:2subfig:acoe}
\begin{minipage}[t]{0.5\linewidth}
\includegraphics[height=1.2in,width=1.5in]{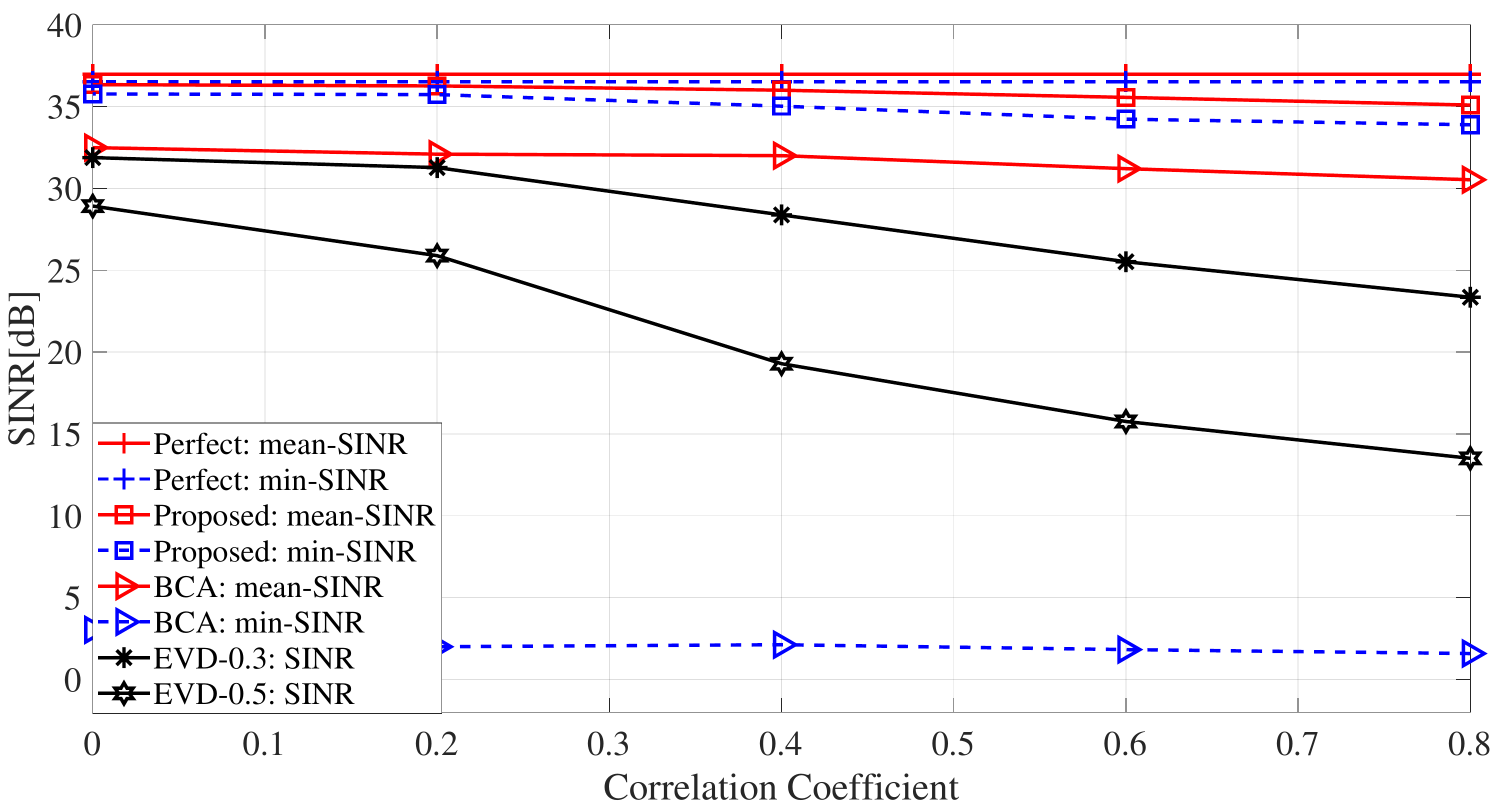}
\end{minipage}
}
\subfigure[]{
\label{fig:2subfig:bcoe}
\begin{minipage}[t]{0.425\linewidth}

\includegraphics[height=1.2in,width=1.5in]{2usercora}
\end{minipage}
}
\centering
\caption{ (a) SINR  and (b)  NMSE  of   proposed,  BCA, and  EVD-based methods versus  correlation coefficient  with $N=1$, $W=1$. The SNR is fixed at 16 dB.}
\label{fig:2usercor}
\end{figure}

\thispagestyle{empty}

\begin{figure}[htbp]
\subfigure[]{
\label{sersnr}
\begin{minipage}[t]{0.5\linewidth}
\includegraphics[height=1.2in,width=1.5in]{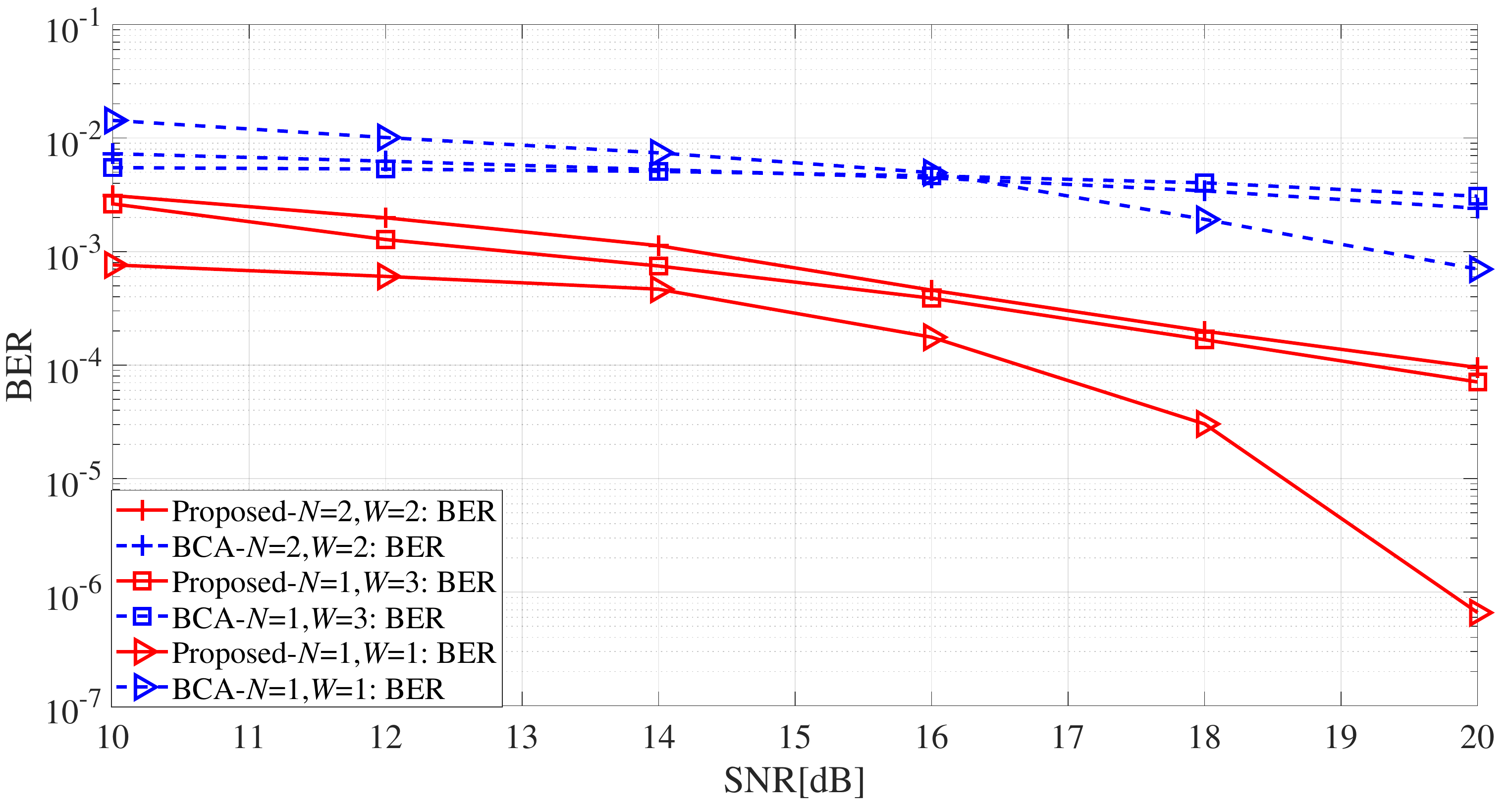}
\end{minipage}
}
\subfigure[]{
\label{sercoe}
\begin{minipage}[t]{0.425\linewidth}
\includegraphics[height=1.2in,width=1.5in]{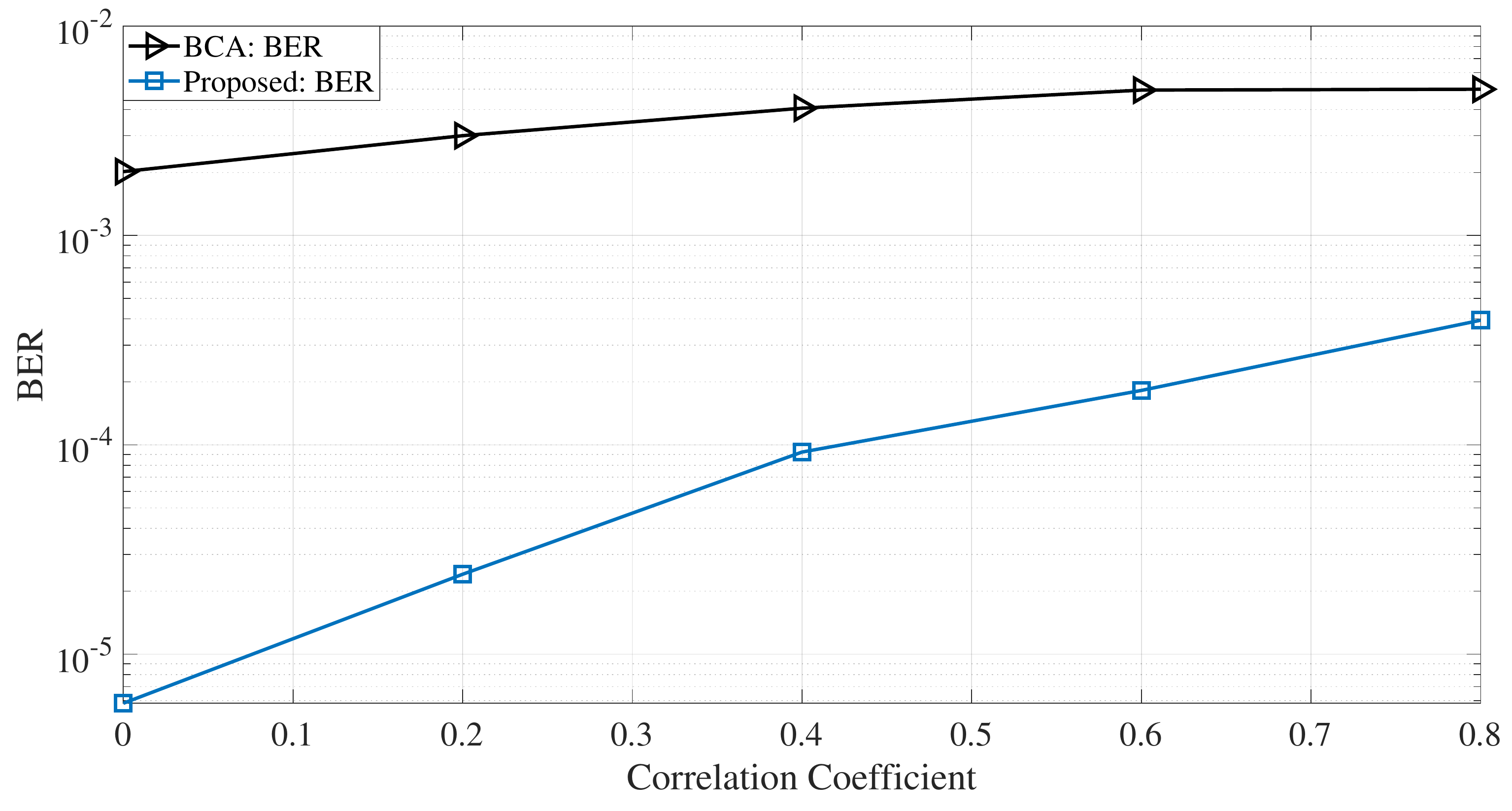}
\end{minipage}
}
\centering
\caption{ BER of proposed  and BCA  methods: (a) $N=2$,  $W=2$; $N=1$,  $W=3$; $N=1$,  $W=1$.  (b) $N=1$,  $W=1$. The SNR is fixed at 16 dB.}
\label{fig:ser}
\end{figure}

\thispagestyle{empty}
{\color{black}
\section{conclusion} \label{con}
\thispagestyle{empty}
Above all, we focus on   channel estimation under a {\color{black}\emph{correlative}} attack with noise.  	We propose an extractor $\mathbb{F}$ that can distill alphabets from {\color{black} a} noisy signal.  {\color{black} We} apply this extractor to   signal  extraction  and channel estimation. Numerical  results show that the proposed method performs
  better  than the BCA  and {\color{black}EVD-based methods} under a \emph{correlative} attack in {\color{black}a} noisy environment.}

\begin{appendices}

\thispagestyle{empty} 
\section{Proof of  correlation coefficient}
When BPSK modulation is used by the signal sequence $\emph{\textbf{a}} \in \mathbb{C}^{1\times n}$, it is not hard to obtain that $P_{A}(1)=P_{A}(-1)=\frac{1}{2}$. $P_{A}$ denotes the stochastic distribution of  $\emph{\textbf{a}}$. 
Due to the IRS, $1\leq t\leq n$, and
\begin{equation}
\emph{\textbf{b}}(t)=\emph{\textbf{a}}(t)e^{\mathsf{i}\phi},
\end{equation}
where   $\phi$ denotes the reflection  phase of the  IRS. It is randomly set according to $\Pr\{\phi=0\}=p$, $\Pr\{\phi=\pi\}=1-p$.
${\emph{\textbf{a}}}(t), {\emph{\textbf{b}}}(t)$ are the respective $t$th elements in ${\emph{\textbf{a}}}, {\emph{\textbf{b}}}$. Then the transition probability $P_{B|A}({1}|{1})=P_{B|A}({-1}|-1)=p$, $P_{B|A}({-1}|{1})=P_{B|A}({1}|{-1})=1-p$. Then in $\emph{\textbf{b}}$, 
\begin{align}
P_{B}(1)&=P_{A}(1)P_{B|A}({1}|{1})+P_{A}(-1)P_{B|A}({1}|{-1}) \\ \nonumber
&=\frac{1}{2}p+\frac{1}{2}(1-p)=\frac{1}{2}.\\
P_{B}(-1)&=1-P_{B}(1)=\frac{1}{2}.
\end{align}

\thispagestyle{empty}

$P_{B}$ denotes the stochastic distribution of  $\emph{\textbf{b}}$. 
 $\overline{\emph{\textbf{a}}}$ and $\overline{\emph{\textbf{b}}}$ are $0$ as  $n$ approaches  infinity, $\overline{(\cdot)}$ denotes the mean value of its input sequence. Then the correlation coefficient of $\emph{\textbf{a}}$ and $\emph{\textbf{b}}$ is 
\begin{align}
\rho _{{\emph{\textbf{a}}}{\emph{\textbf{b}}}}&=\frac{\sum_{t=1}^{n}(\emph{\textbf{a}}(t)-\overline{\emph{\textbf{a}}})(\emph{\textbf{b}}(t)-\overline{\emph{\textbf{b}}})}
{\sqrt{\sum_{t=1}^{n}(\emph{\textbf{a}}(t)-\overline{\emph{\textbf{a}}})^2}\sqrt{\sum_{t=1}^{n}(\emph{\textbf{b}}(t)-\overline{\emph{\textbf{b}}})^2}}\\ \nonumber
&=\frac{\sum_{t=1}^{n}\emph{\textbf{a}}(t)\emph{\textbf{b}}(t)}
{\sqrt{\sum_{t=1}^{n}\emph{\textbf{a}}(t)^2}\sqrt{\sum_{t=1}^{n}\emph{\textbf{b}}(t)^2}}=\frac{pn-(1-p)n}{n}\\ \nonumber
&=2p-1.
\end{align}

\thispagestyle{empty}

\thispagestyle{empty}
\section{Proof of Proposition 1}
According to [\cite{ee}, Corollary 1], %for $j=1, \cdots ,N$ and $j=1, \cdots, N$. 
we obtain $\frac{1}{M}\textbf{\emph{H}}^H\textbf{\emph{H}}\xrightarrow{a.s.}\frac{1}{M}{\rm{Tr(R_H)}}, \frac{1}{M}\textbf{\emph{G}}^H\textbf{\emph{G}}\xrightarrow{a.s.}\frac{1}{M}{\rm{Tr(R_G)}}, \frac{1}{M}\textbf{\emph{G}}^H\textbf{\emph{H}}\xrightarrow{a.s.}0, \frac{1}{M}\textbf{\emph{H}}^H\textbf{\emph{G}}\xrightarrow{a.s.}0$.
\thispagestyle{empty}
\thispagestyle{empty}Then we have 
{ $\frac{1}{M}\textbf{\emph{C}}^H\textbf{\emph{C}}\xrightarrow{a.s.} \frac{1}{M} \left[ \begin{matrix} \rm{Tr(R_H)} &0\\ 0&\rm{Tr(R_G)}\end{matrix}\right] =R$}, and $U^H_{\textbf{\emph{Y}}}U_{\textbf{\emph{Y}}}\xrightarrow{a.s.}\emph{\textbf{I}}_M$, %Because the distribution of $\textbf{\emph{S}}$ is independent and identical, 
and we have $\frac{\textbf{\emph{S}}\textbf{\emph{S}}^H}{n}\xrightarrow{a.s.}{{R_s}}$. 
Then we decompose $R_\emph{\textbf{{Y}}}$  %$\frac{1}{Mn}\emph{\textbf{{Y}}}\emph{\textbf{{Y}}}^H$ 
as 
%\begin{footnotesize}
\begin{align}
R_\emph{\textbf{{Y}}}&=\frac{1}{M}{\textbf{\emph{C}}}R_s\textbf{\emph{C}}^H+\frac{\sigma^2}{M}\emph{\textbf{I}}_M \\ \nonumber
&=\frac{1}{M}\textbf{\emph{C}}R^{-\frac{1}{2}}R^{\frac{1}{2}}R_sR^{\frac{1}{2}}R^{-\frac{1}{2}}\textbf{\emph{C}}^H+\frac{\sigma^2}{M}\emph{\textbf{I}}_M \\ \nonumber
& \overset{(a)}{=}\frac{1}{M}\textbf{\emph{C}}R^{-\frac{1}{2}} U_s \Lambda  U^H_s R^{-\frac{1}{2}}\textbf{\emph{C}}^H+\frac{\sigma^2}{M}\emph{\textbf{I}}_M \\ \nonumber
&\xrightarrow{a.s.} \frac{1}{M}\textbf{\emph{C}}R^{-\frac{1}{2}} U_s \Lambda  U^H_s R^{-\frac{1}{2}}\textbf{\emph{C}}^H+\frac{\sigma^2}{M}U_{\emph{\textbf{{Y}}}}U^H_{\emph{\textbf{{Y}}}} \\ \nonumber
&=\frac{1}{M}\textbf{\emph{C}}R^{-\frac{1}{2}} U_s \Lambda  U^H_s R^{-\frac{1}{2}}\textbf{\emph{C}}^H+\frac{\sigma^2}{M}U_{W}U^H_{W}+\\& \nonumber \frac{\sigma^2}{M}\textbf{\emph{C}}R^{-\frac{1}{2}} U_s  U^H_s R^{-\frac{1}{2}}\textbf{\emph{C}}^H  \\ \nonumber
&=U_{\textbf{\emph{Y}}}diag\{\sigma^{2} \textbf{I}_{M-N-WN}, \Lambda + \sigma^{2}\textbf{I}_{N+WN} \}U^H_{\textbf{\emph{Y}}}.
\end{align}
%\end{footnotesize}
As long as $n$ is sufficiently large, $R_\emph{\textbf{{Y}}}$ can be approached by $\frac{1}{Mn}\emph{\textbf{{Y}}}\emph{\textbf{{Y}}}^H$. 
The convergence  follows $U_{\textbf{\emph{Y}}}U^H_{\textbf{\emph{Y}}}\xrightarrow{a.s.}\emph{\textbf{I}}_M$, and  the equation ($a$) follows  $R^{\frac{1}{2}}R_sR^{\frac{1}{2}}=U_s\Lambda U^H_s$.

\section{Proof of Proposition 2}

Due to space limitations, we sketch the proof of Proposition 2 and present
an algorithm to implement it. 
We use $L^n$ to denote $V^n+Q^n$. 
Note that $L^n$, $V^n$, $Q^n$
are i.i.d. random sequences.
Let  $V$ denote a generic random variable with
the same stochastic distribution as   each element of $V^n$.
Similarly, we  use  generic random variables $Q$ and $L$
following stochastic distributions identical to those of elements of $Q^n$ and $L^n$, respectively.
Furthermore, note that $Q^n$ and $V^n$ are independent of each other. As a consequence,
 $W$ is independent of $V$.
Since $L^n=V^n+Q^n$, we  specify $L$ by $L=V+Q$.
 Let $F_{{L}}$,
$F_{{V}}$, and $F_{Q}$ denote the distributions of
$L$, $V$, and $Q$,
respectively. Then, because $V$ and $Q$ are
independent of each other, we have
\begin{equation} \label{eq:chi}
\Phi_{F_{L}}
  (\boldsymbol{f}) = \Phi_{F_{V}}(\boldsymbol{f})
 \Phi_{F_{Q}}(\boldsymbol{f}),
\end{equation}where $\Phi_F( \boldsymbol{f})$ denotes the characteristic
function (CF) of the distribution $F$, and
$\boldsymbol{f}$ is the 
frequency vector. Note that the noise variance parameter $\sigma_{Q}^2$ is
a characteristic of the receiver circuitry and can be measured
\emph{a priori}. We may assume that its value is known; hence,
$\Phi_{F_{Q}}(\boldsymbol{f})
  =\exp\left\{ -2{\sigma_{Q}^{2}}\pi^2\left| \boldsymbol{f}\right|^{2}\right\}$
is also known. Therefore, according to (\ref{eq:chi}), $F_{V}$ is achieved by

\begin{equation}\label{esti_ideal}
F_{V}=\Phi^{-1}\left(\frac{\Phi_{F_{L}}(\boldsymbol{f})}{\exp\left\{ -2{\sigma_{Q}^{2}}\pi^2\left|\boldsymbol{f}\right|^{2}\right\} }\right),
\end{equation}where $\Phi^{-1}\left(\cdot\right)$ denotes the inverse CF of its input.
 \thispagestyle{empty}It is worth noting that in (\ref{esti_ideal}), $F_{V}$ is perfectly obtained from $L$, even though $L$ includes noise $Q$ with arbitrary average
power $\sigma_{Q}^{2}$.   $V$ has a discrete alphabet $\mathcal{V}$ that can be achieved by finding points $\mathsf{v}$
that make $F_{V}\left(\mathsf{v}\right)>0$. As a result, the extractor given by (\ref{esti_ideal}) satisfies our goal of extracting alphabets from noisy observations.
However, in practice, to implement   (\ref{esti_ideal}) is a challenge for two reasons.
\begin{enumerate}
\item Due to attack, $F_{L}$ is unknown. The lack of $F_{L}$ leads the inability to obtain $\Phi_{F_{L}}(\boldsymbol{\omega})$ exactly.
\item $\Phi\left(\cdot\right)$ and $\Phi^{-1}\left(\cdot\right)$ correspond to a continuous Fourier transform (CFT) and inverse CFT, respectively.
The transforms over a continuous domain may give rise to issues of implementation.
\end{enumerate}
 \thispagestyle{empty} 
Motivated by these two challenges, we propose an extractor according to (\ref{esti_ideal}) by using a quantized empirical distribution of $L^n$ to approach $F_{L}$ according to the law of large numbers (LLN), and using a discrete Fourier transform (DFT) and inverse DFT to approach $\Phi\left(\cdot\right)$ and $\Phi^{-1}\left(\cdot\right)$, respectively. 
According to LLN and the Nyquist sampling theorem, the approximation of $\Phi_{F_{L}}(\boldsymbol{f})$ becomes more accurate as the quantization level and number of observations increase. {In this sense, on the basis of (\ref{esti_ideal}), Proposition 2 has been proved.}
Furthermore, we provide an algorithm to implement (\ref{esti_ideal}) by sequential quadratic programming (SQP). 
To be more precise,  notice that (\ref{esti_ideal}) is equivalent to 
{\small{\begin{equation}\label{opt}
F_{V}\left(v\right)=\underset{\hat{F}_{V}}{\arg\min}\iint\left|\Phi_{F_{L}}(\boldsymbol{f})-\Phi_{\hat{F}_{V}}(\boldsymbol{f})\Phi_{F_{Q}}(\boldsymbol{f})\right|^{2}d\boldsymbol{f},
\end{equation}}}where $\hat{F}_{V}$ is a stochastic distribution function. 
To approximate $\Phi_{F_{L}}(\boldsymbol{f})$, we quantize $L$ and achieve an empirical distribution, 
{\small{\begin{equation}\label{empirical_dis}
\varPi_{\widehat{L}^{n}}\left(n_{1},n_{2}\right)=\frac{1}{n}\sum_{i=1}^{n}\mathtt{1}\left\{ \Re\left\{ L_{i}\right\} \in\mathcal{B}\left(n_{1}\right)\right\} \mathtt{1}\left\{ \Im\left\{ L_{i}\right\} \in\mathcal{B}\left(n_{2}\right)\right\} ,
\end{equation}}}where $L_{i}$ is the $i$-th variable of $L^n$;  $\Re\left\{ \cdot\right\}$ and   $\Im\left\{ \cdot\right\} $ denote the real and  imaginary parts, respectively, of its input; and
$\mathtt{1}\left\{ \cdot\right\} $ is an indicator function. $\mathcal{B}\left(n_{1}\right)=\left[-d_{1}+n_{1}\triangle,-d_{1}+\left(n_{1}+1\right)\triangle\right]$, 
$\triangle=\frac{2d_{1}}{n_{1}}$, $d_{1}=\sqrt{n_{1}}$. For $\boldsymbol{f}=\left[f_{r},\: f_{i}\right]$, $\Phi_{F_{L}}(\boldsymbol{f})$ could be approached by 
{\small{\begin{align}
&\Phi_{F_{L}}(\boldsymbol{f})=\int F_{L}\left({l}\right)\exp\left\{ -\mathsf{i}2\pi\boldsymbol{f}\left[\begin{array}{c}
\Re\left({l}\right)\\
\Im\left({l}\right)
\end{array}\right]\right\} d{l}\\&\nonumber\overset{N,n\rightarrow\infty}{\rightarrow}\triangle^{2}\exp\left\{ \mathsf{i}2\pi\left(f_{r}+f_{i}\right)d_{1}\right\}\\&\nonumber\times\sum_{n_{1}=1}^{N}\sum_{n_{2}=1}^{N}\varPi_{\widehat{L}^{n}}\left(n_{1},n_{2}\right)\exp\left\{ -\mathsf{i}2\pi n_{1}\triangle f_{r}\right\}\exp\left\{ -\mathsf{i}2\pi n_{2}\triangle f_{i}\right\}.
\end{align}}}Sampling $\Phi_{F_{L}}(\boldsymbol{f})$ across $\left(k_{1}f,\: k_{2}f\right)$, $k_{1},k_{2}=1,\ldots,N_{f}$, $f=\frac{1}{\triangle N_{f}}$, we have 
{\small{\begin{align}
&\Phi_{F_{L}}(k_{1}f,\: k_{2}f)\nonumber\overset{N,n\rightarrow\infty}{\rightarrow}\tilde{\Phi}_{F_{L}}(k_{1}f,\: k_{2}f)\\\nonumber&=\triangle^{2}\exp\left\{ \mathsf{i}2\pi\left(\frac{k_{1}}{\triangle N_{f}}+\frac{k_{2}}{\triangle N_{f}}\right)d_{1}\right\}\\&\nonumber\times\underbrace{\sum_{n_{1}=1}^{N}\sum_{n_{2}=1}^{N}\varPi_{\widehat{L}^{n}}\left(n_{1},n_{2}\right)\exp\left\{ -\mathsf{i}2\pi\frac{n_{1}k_{1}}{N_{f}}\right\} \exp\left\{ -\mathsf{i}2\pi\frac{n_{2}k_{1}}{N_{f}}\right\} }_{\left[\mathbb{DFT}\left\{ \varPi_{\widehat{L}^{n}}\right\} \right]_{k_{1},k_{2}}},
\end{align}}}where $\tilde{\Phi}_{F_{L}}(k_{1}f,\: k_{2}f)$  can be obtained from the DFT of $\varPi_{\widehat{L}^{n}}$, denoted by $\mathbb{DFT}\left\{ \varPi_{\widehat{L}^{n}}\right\} $,
which is an $N_f\times N_f$ matrix whose $(k_1, k_2)$-th element corresponds to the value of $\mathbb{DFT}\left\{ \varPi_{\widehat{L}^{n}}\right\} $ in the $(k_1, k_2)$-th frequency. Hence, we approximate $\Phi_{F_{L}}(\boldsymbol{f})$ by an $N_f\times N_f$ matrix $\mathbf{L}$ whose $(k_1, k_2)$-th element is 
{\small\begin{equation}\label{L_matrix}
\left[\mathbf{L}\right]_{k_{1},k_{2}}=\triangle^{2}\exp\left\{ \mathsf{i}2\pi\left(\frac{k_{1}}{\triangle N_{f}}+\frac{k_{2}}{\triangle N_{f}}\right)d_{1}\right\} \left[\mathbb{DFT}\left\{ \varPi_{\widehat{L}^{n}}\right\} \right]_{k_{1},k_{2}}.
\end{equation}}
Similarly, $\Phi_{{F}_{V}}(\boldsymbol{f})$ can be approximated by an $N_f\times N_f$ matrix $\mathbf{V}$ whose $(k_1, k_2)$-th element is 
{\small\begin{equation}\label{V_ele}
\left[\mathbf{V}\right]_{k_{1},k_{2}}=\triangle^{2}\exp\left\{ \mathsf{i}2\pi\left(\frac{k_{1}}{\triangle N_{f}}+\frac{k_{2}}{\triangle N_{f}}\right)d_{1}\right\} \left[\mathbb{DFT}\left\{ \varPi_{\widehat{V}^{n}}\right\} \right]_{k_{1},k_{2}}.
\end{equation}}$\varPi_{\widehat{V}^{n}}$ is the empirical distribution of the quantized sequence of $V^n$, similar to $\varPi_{\widehat{L}^{n}}$ (\ref{empirical_dis}). 
Furthermore, according to the definition of DFT, we extend $\mathbb{DFT}\left\{ \varPi_{\widehat{V}^{n}}\right\} $ by

\begin{small}
\begin{equation}\label{DFT}
\mathbb{DFT}\left\{ \varPi_{\widehat{V}^{n}}\right\} =\mathbf{F}\varPi_{\widehat{V}^{n}}\mathbf{F}^{T},
\end{equation}
\end{small}where $\mathbf{F}\in\mathbb{C}^{N_{f}\times N}$, $\left[\mathbf{F}\right]_{i,j}= \exp\left(-j2\pi (i-1)(j-1)/N_{f}\right)$, $i=1,\cdots,N$, $j=1,2,\cdots,N_{f}$. 
Substituting (\ref{DFT}) in (\ref{V_ele}), we have
\begin{equation}\label{V_matrix}
\mathbf{V}=\mathbf{R}_{V}\odot\left\{ \mathbf{F}\varPi_{\widehat{V}^{n}}\mathbf{F}^{T}\right\},
\end{equation}where $\mathbf{R}_{V}\in\mathbb{C}^{N_{f}\times {N_f}}$, $\left[\mathbf{R}_{V}\right]_{k_{1},k_{2}}=\triangle^{2}\exp\left\{ \mathsf{i}2\pi\left(\frac{k_{1}}{\triangle N_{f}}+\frac{k_{2}}{\triangle N_{f}}\right)d_{1}\right\}$, $k_1,k_2=1,\ldots, N_f$, and $\odot$ denotes the dot product. 
Notice that $\Phi_{F_{L}}(\boldsymbol{f})$ and $\Phi_{F_{V}}(\boldsymbol{f})$ can be approximated by $\mathbf{L}$ and $\mathbf{V}$, respectively. 
Based on (\ref{eq:chi}), we have 
{\small{\begin{equation}\label{chi_matrix}
\mathbf{L}\approx\mathbf{R}_{Q}\odot\mathbf{R}_{V}\odot\left\{ \mathbf{F}\varPi_{\widehat{V}^{n}}\mathbf{F}^{T}\right\},
\end{equation}}}where $\mathbf{R}_{Q}\in\mathbb{C}^{N_{f}\times {N_f}}$ samples $\Phi_{F_{Q}}(\boldsymbol{f})$, $\left[\mathbf{R}_{Q}\right]_{k_{1},k_{2}}=\exp\left\{ -2{\sigma_{Q}^{2}}\pi^{2}\left|k_{1}^{2}+k_{2}^{2}\right|f^{2}\right\}$. 
Then (\ref{eq:chi}) further indicates that (\ref{opt}) can be transformed to a matrix form, 
{\small{\begin{equation}\label{opt_alter}
\widetilde{F}_{V}=\underset{\hat{F}_{V}}{\arg\min}\left|\mathbf{L}-\mathbf{R}_{Q}\odot\mathbf{R}_{V}\odot\left\{ \mathbf{F}\hat{F}_{V}\mathbf{F}^{T}\right\} \right|^{2},
\end{equation}}}where $\hat{F}_{V}\in\mathbb{C}^{N\times N}$ is a stochastic matrix to characterize the distribution over a complex domain.
We use SQP to solve (\ref{opt_alter}). The points making $\widetilde{F}_{V}$ achieve local maxima are extracted as the estimate of $\mathcal{V}$.
As $n$, $N$, and $N_f$ increase, (\ref{opt_alter}) approximates (\ref{opt}) more accurately. As a beneficial result, the extracted points from $\widetilde{F}_{V}$
converge to  $\mathcal{V}$ in probability.
We define the proposed extractor as $ \mathbb{{F}}$, whose steps are summarized by {Algorithm 2}, which can be run several times to achieve  convergence.  
\begin{algorithm}[h]
  \caption{ $\mathbb{F}$: Extraction of Alphabets from a Noisy Sequence $L^n$}
  \label{alg::conjugateGradient}
  \begin{algorithmic}[1]
	\State Get $\varPi_{\widehat{L}^{n}}$ from $L^n$ according to (\ref{empirical_dis})
        \State Get $\mathbf{L}$ according to (\ref{L_matrix})
   \State Set up optimization problem (\ref{opt_alter})
{\small{$\widetilde{F}_{V}=\underset{\hat{F}_{V}}{\arg\min}\left|\mathbf{L}-\mathbf{R}_{Q}\odot\mathbf{R}_{V}\odot\left\{ \mathbf{F}\hat{F}_{V}\mathbf{F}^{T}\right\} \right|^{2}$}}
  \State  Invoking SQP method to solve (\ref{opt_alter})
\State  Based on $\widetilde{F}_{V}$,  find the local maximum points for extracting alphabet. 
  \end{algorithmic}
\end{algorithm}

\section{Proof of Proposition 3}
	We notice that {\color{black}the optimized signal  extraction vector}  $\hat{\emph{\textbf{u}}}$ only {\color{black}extracts the signal of one user, $\textbf{\emph{s}}$. Further, estimate the corresponding channel $\textbf{\emph{c}}$. Then we rewrite  $\textbf{\emph{Y}}$ as}% but we don't know the signal corresponds to which user clearly,  we assume that the extracted signal  is $a_j$, which represents the message sequence of the $j$-th LUs, and the channel is $\emph{\textbf{h}}_j$. 
%%%% The (\ref{matrixform}) can be transformed into:
% First, after optimization, we get $\hat{\emph{\textbf{u}}}$
%\begin{footnotesize}
\begin{equation} \label{al}
\begin{aligned}
 \textbf{\emph{Y}}=\textbf{\emph{R}}+\textbf{\emph{c}}\textbf{\emph{s}},
%S=\hat{\emph{\textbf{u}}}\emph{\textbf{Y}}&=\hat{\emph{\textbf{u}}}\emph{\textbf{h}}_j\sqrt{P_A}a_j +\\  &\hat{\emph{\textbf{u}}} \Bigg\{ \sum_{j' \neq j}^{N_L}{\emph{\textbf{h}}_j'}\sqrt{P_A}a_j' +\sum_{k=1}^{N_M} {\emph{\textbf{g}}_k}\sqrt{P_B}b_k \Bigg\} + \hat{\emph{\textbf{u}}}\textbf{\emph{N}}
%\emph{\textbf{Y}}=\emph{\textbf{h}}_j\sqrt{P_A}a_j + \Bigg\{ \sum_{j' \neq j}^{N_L}{\emph{\textbf{h}}_j'}\sqrt{P_A}a_j' +\sum_{k=1}^{N_M} {\emph{\textbf{g}}_k}\sqrt{P_B}b_k \Bigg\} +\textbf{\emph{N}}
\end{aligned}
\end{equation}where $\textbf{\emph{R}}$ is the remainder signal.\thispagestyle{empty}   More precisely, we choose the $m$th row of \emph{\textbf{Y}}, where $[\cdot]_m$  denotes the  $m$th row of its input matrix or vector,
\begin{equation} \label{alphabets}
[\textbf{\emph{Y}}]_m=[\textbf{\emph{R}}]_m+[\textbf{\emph{c}}]_m\textbf{\emph{s}}.
%[\emph{\textbf{Y}}]_m=[\emph{\textbf{h}}_j]_m\sqrt{P_A}a_j + r + [\textbf{\emph{N}}]_m
%{\footnotesize$r=\Bigg\{ \sum_{j' \neq j}^{N_L}{[\emph{\textbf{h}}_j']_m}\sqrt{P_A}a_j' +\sum_{k=1}^{N_M} {[\emph{\textbf{g}}_k}\sqrt{P_B}b_k ]_m\Bigg\}$}. 
%If there is noiseless signal, then 
\end{equation}Since the noise exists, we use the extractor $\mathbb{F}$ to distill alphabets and obtain the  alphabets of  (\ref{alphabets}) as
\begin{equation}
\mathcal{Y}_m=\left\{ \mathsf{y}| \mathsf{y} =\gamma+[\textbf{\emph{c}}]_m \mathsf{z},    \mathsf{z}\in \mathcal{Z}, \gamma \in\mathcal{R}\right\},
\end{equation}where $\mathcal{Y}_m=\mathbb{F}\left\{ [\textbf{\emph{Y}}]_m\right\}$, $\mathcal{Z}=\mathbb{F}\left\{ \textbf{\emph{s}}\right\}$, and $\mathcal{R}=\mathbb{F}\left\{[\textbf{\emph{R}}]_m \right\}$. We discover that all   the different pairwise elements $(\mathsf{y}-\mathsf{y}')$  chosen from  $\mathcal{Y}_m$ must contain the element   $[\textbf{\emph{c}}]_m (\mathsf{z}-\mathsf{z}')$. %in \cite{cp}. 
Finally, we can obtain the finite set of $[\textbf{\emph{c}}]_m$ as
%\begin{footnotesize}
 \thispagestyle{empty} 
\begin{equation}
\begin{aligned}
\mathrm{Q_{m}}=\left\{ q\:|\: q=\frac{\mathsf{y}-\mathsf{y}'}{\mathsf{z}-\mathsf{z}'},\mathsf{z}\neq\mathsf{z}',\mathsf{y}\neq\mathsf{y}',  \mathsf{y},\mathsf{y}'\in\mathcal{Y}_m,  \mathsf{z},\mathsf{z}'\in \mathcal{Z}\right\}.
\end{aligned}
 \end{equation}

\end{appendices}

\thispagestyle{empty}
\bibliographystyle{ieeetr}
\bibliography{this}
\thispagestyle{empty}
 \end{document}